\documentclass[sigconf]{acmart}
\usepackage{subcaption}
\usepackage{makecell}
\usepackage{multirow}
\usepackage{multicol}
\usepackage{enumitem}
\usepackage{lipsum}
\usepackage{mathtools}
\usepackage{cuted}
\usepackage{xcolor}

\DeclareMathOperator*{\argmax}{argmax}
\DeclareMathOperator*{\argmin}{argmin}
\AtBeginDocument{%
  \providecommand\BibTeX{{%
    \normalfont B\kern-0.5em{\scshape i\kern-0.25em b}\kern-0.8em\TeX}}}

\copyrightyear{2019}
\acmYear{2019}
\setcopyright{rightsretained}

\acmConference{}
\acmDOI{}
\acmISBN{}
\acmBooktitle{}

\begin{document}

\newcommand{\cikm}[1]{{#1}}

\title{Noise Contrastive Estimation for Autoencoding-based\\ One-Class Collaborative Filtering}


\author{Jin Peng Zhou}
\affiliation{%
  \institution{University of Toronto}
}
\email{jinpeng.zhou@mail.utoronto.ca}

\author{Ga Wu}
\affiliation{%
  \institution{University of Toronto}
}
\email{wuga@mie.utoronto.ca}

\author{Zheda Mai}
\affiliation{%
  \institution{University of Toronto}
}
\email{zheda.mai@mail.utoronto.ca}

\author{Scott Sanner}
\affiliation{%
  \institution{University of Toronto}
}
\email{ssanner@mie.utoronto.ca}

\newcommand{\norm}[1]{\left\lVert#1\right\rVert}

\begin{abstract}
    One-class collaborative filtering (OC-CF) is a common class of recommendation problem where only the positive class is explicitly observed (e.g., purchases, clicks). 
    Autoencoder based recommenders such as AutoRec and variants demonstrate strong performance on many OC-CF benchmarks, but also empirically suffer from a strong popularity bias.
    While a careful choice of negative samples in the OC-CF setting can mitigate popularity bias, Negative Sampling (NS) is often better for training embeddings than for the end task itself. To address this, we propose a two-headed AutoRec to first train an embedding layer via one head using Negative Sampling then to train for the final task via the second head. While this NS-AutoRec improves results for AutoRec and outperforms many state-of-the-art baselines on OC-CF problems, we notice that Negative Sampling can still take a large amount of time to train. Since Negative Sampling is known to be a special case of Noise Contrastive Estimation (NCE), we adapt a recently proposed closed-form NCE solution for collaborative filtering to AutoRec yielding NCE-AutoRec. 
    \cikm{Overall, we show that our novel two-headed  AutoRec models (NCE-AutoRec and NS-AutoRec) successfully mitigate the popularity bias issue and maintain competitive performance in comparison to state-of-the-art recommenders on multiple real-world datasets.}
\end{abstract}

% \keywords{One-Class Collaborative Filtering, Noise Contrastive Estimation, Autoencoder, Recommender System}


\maketitle

\section{Introduction}
Recommender systems are an essential part of e-commerce platforms that help users discover items of interest ranging from movies to restaurants. Collaborative filtering (CF) is the de facto standard approach for making these personalized recommendations based on user-item interaction data from a general population~\cite{sarwar2002recommender}. In many practical recommendation settings, interactions such as clicks on a website or purchase history are defined as positive-only signals while explicit negative signals are missing. In such settings, it is critical to avoid treating the absence of interactions as implicit negative signals since they may simply be attributed to a user's unawareness of an item. We refer to the setting where only positive interactions (or preferences) are observed as the One-Class Collaborative Filtering (OC-CF) problem~\cite{pan2008one}.

One effective approach to OC-CF problems is via Autoencoding-based methods such as AutoRec~\cite{Sedhain:2015:AAM:2740908.2742726}, which aim to learn non-linear latent features by first encoding interactions into a lower-dimensional latent representation and then decoding them to approximately reconstruct the originally observed interactions while also providing predictions for unobserved interactions. By employing a nonlinear activation function in the encoding layer, AutoRec is able to represent a more complex relationship between users and items than simpler linear embedding approaches such as probabilistic matrix factorization~\cite{Salakhutdinov:2007:PMF:2981562.2981720} or PureSVD~\cite{cremonesi2010performance}.  Recently, a growing community of researchers applying the Autoencoder framework to collaborative filtering have demonstrated promising results~\cite{liang2018variational, wu2016collaborative, strub2015collaborative, strub2016hybrid}.

However, as we will demonstrate empirically in this work, the distributions of items recommended by AutoRec and its variants exhibit a strong bias towards more popular items. Motivated by recent work that mitigated popularity bias through Negative Sampling and its closed-form Noise Contrastive Estimation (NCE) generalization~\cite{sigir19a}, we consider in this work whether such methods can be used to improve Autoencoder based methods.

To this end, we propose two novel two-headed Autoencoder based methods, Negative Sampling enhanced AutoRec (NS-AutoRec) and Noise Contrastive Estimation enhanced AutoRec (NCE-AutoRec). Instead of treating unobserved interactions as an explicit negative signal, we leverage insights from Negative Sampling (NS) that have been extremely effective in learning word embeddings~\cite{mikolov2013distributed}.
NS-AutoRec first trains a self-supervised embedding layer via one head by Negative Sampling and then trains for the end prediction task via the second head. We notice, however, that the training time of NS-AutoRec increases with the number of negative samples --- impeding scalability on particularly large datasets. 

Recent research \cite{bose-etal-2018-adversarial, rao2016noise} shows Negative Sampling is a special case of Noise Contrastive Estimation~(NCE)~\cite{gutmann2010noise}, hence we adapt this more efficient closed-form NCE solution to AutoRec yielding NCE-AutoRec. Specifically, we first learn item embeddings by training with a ``de-popularized'' data transformation as the objective of an NCE head. Because NCE itself is not ideal for the final prediction task, we then train a separate decoder head that uses the learned item embeddings to perform the end task. NCE therefore plays two critical roles in this method: (i) ``de-popularizing'' the item embeddings as in NS-AutoRec and (ii) accelerating the embedding training compared to NS by sidestepping the need to sample.

In summary, this paper makes two major key contributions:
% to Autoencoder based collaborative filtering in the OC-CF setting:
\begin{itemize}
    \item We propose two variants of a novel two-headed Autoencoder based recommendation method called NS-AutoRec and NCE-AutoRec that respectively employ negative sampling and an alternative closed-form NCE solution %with standard MSE optimization objectives 
    to learn high quality ``de-popularized'' embeddings \emph{without} sacrificing end task loss. Empirical results show that NCE-AutoRec is more efficient and scalable than NS-AutoRec.
    
    % \item We propose a scalable two-headed Autoencoder based recommendation method called NCE-AutoRec that employs both NCE and standard MSE optimization objectives to learn high quality embeddings without sacrificing end task loss.
    
    \item We demonstrate that both NS-AutoRec and NCE-AutoRec are 
    %able to outperform 
    \cikm{competitive with}
    many state-of-the-art recommendation methods on four large-scale publicly available datasets while 
    \cikm{providing more personalization (i.e., focusing on recommending less popular and thus more nuanced items)}.
\end{itemize}

Hence, our novel AutoRec extensions exhibit high personalization while maintaining state-of-the-art OC-CF performance.
%two-headed NCE extension of AutoRec provides critical improvements for the practically important OC-CF setting.

\section{Notation and Related Work}
In this section, we briefly review Autoencoder based recommender systems such as AutoRec~\cite{Sedhain:2015:AAM:2740908.2742726}. Then we describe Negative Sampling~\cite{mikolov2013distributed} based ranking losses that are widely used in modern recommender systems. Finally, we discuss theoretical results~\cite{bose-etal-2018-adversarial, rao2016noise} that establish Negative Sampling as a Monte Carlo approximation of Noise Contrastive Estimation~\cite{gutmann2010noise} that we leverage in this work.

\begin{figure*}[thp!]
  \centering
  \includegraphics[width=0.7\linewidth]{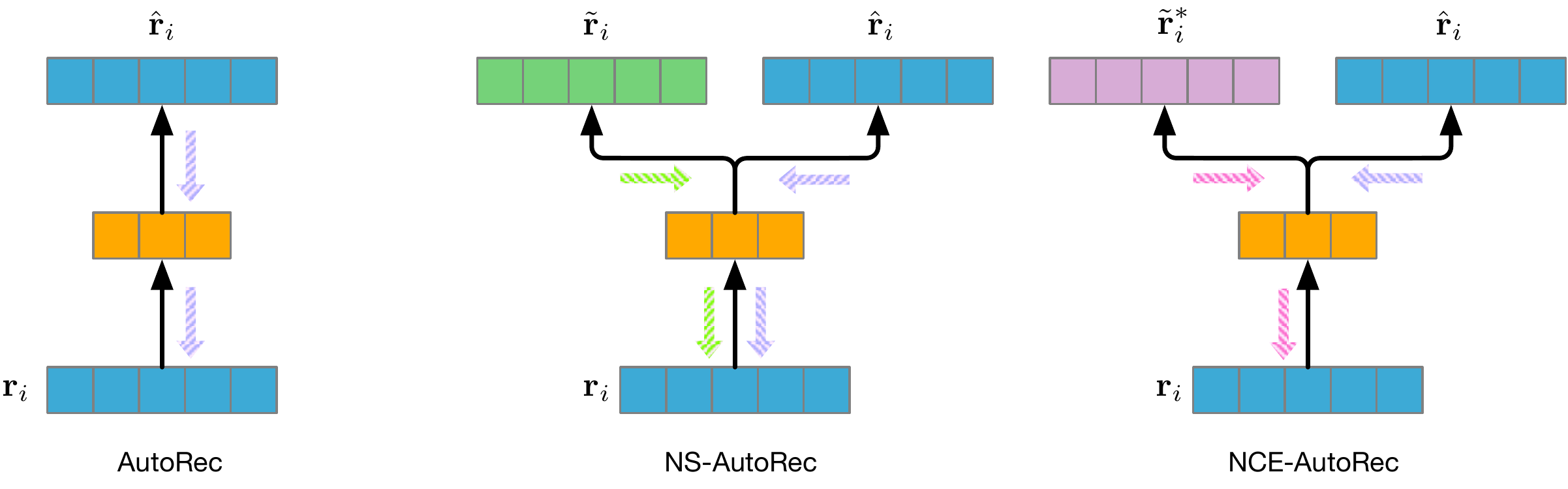}
  \caption{Architectures of the proposed recommender systems NS-AutoRec and NCE-AutoRec. (left) The original AutoRec architecture that reproduces its input through simple forward propagation. (middle) Negative Sampling enhanced AutoRec that jointly optimizes both NS (green) and MSE (blue) objectives. (right) Noise Contrastive Estimation enhanced AutoRec that learns an encoder network through optimizing the NCE objective (pink) and learns the decoder network through optimizing the MSE objective. The dashed arrows show backpropagation flows.}
  \label{fig:nce-autorec}
\end{figure*}

\subsection{Notation}
Before proceeding, we define our notation as follows:
\begin{itemize}
% Don't switch terminology on the reader -- you have not mentioned "implicit" before
    \item $R$: The positive-only feedback matrix for $m$ users and $n$ items in the shape of $m \times n$. Each entry of $r_{i, j}$ is either 1, which indicates there is an interaction between user $i$ and item $j$, or 0 otherwise (no interaction). We use $r_{:, j}$ to represent all user feedback for item $j \in \{ 1 \cdots n \}$.
    % Wuga commited below as norm-1 is clear enough
    %and $||r_{:, j}||_1$ to represent the number of observed interactions for item $j$.
    \item $\hat{R}$ and $\Tilde{R}$: Respective reconstructions of the original matrix $R$ based on Mean Squared Error~(MSE) and Negative Sampling~(NS) (in NS-AutoRec) or closed-form NCE solution (in NCE-AutoRec). Both have the same matrix shape as $R$.
    \item $R^*$: The matrix that optimizes the NCE objective. It has the same shape as the matrix $R$. We will describe the NCE objective function shortly.
    % \item $U, V, \textit{and } W$: Weights of fully connected layers in a neural network. The bias term associated with network are correspondingly defined as $b_{U}$, $b_{V}$ and $b_{W}$.
\end{itemize}

\subsection{AutoRec}
AutoRec~\cite{Sedhain:2015:AAM:2740908.2742726} is an Autoencoder based recommender system that aims to encode the historical interactions of each user into a low-dimensional user representation (a vector), and then reconstruct the observations by decoding the learned latent representation. AutoRec makes predictions for all items by leveraging generalization from its reduced dimensional latent user embedding. 

AutoRec aims to minimize a Mean Squared Error~(MSE) objective function
\begin{equation}
\sum_{i}\norm{\mathbf{r}_{i} - \hat{\mathbf{r}}_{i}}^{2}_2 + \lambda \norm{\Omega}^{2}_{F}
\label{eq:autorec-loss}
\end{equation}
where we use $\Omega$ to represent all parameters in the AutoRec model. The prediction vector $\hat{\mathbf{r}}_i$ is produced through forward propagation of the AutoRec network architecture as shown in Figure \ref{fig:nce-autorec} (left).

The original AutoRec was proposed to tackle recommendation with explicit feedback, where the goal is to reconstruct numerical ratings (e.g., the MovieLens dataset has ordinal ratings $\{1,2,3,4,5\}$)
%\footnote{Rating is a score that the user provides in a certain range. For the MovieLens dataset, the ratings are from 1 to 5.} 
from historical observations.
%, which is not well-suited for top-K ranking tasks with positive-only feedback. 
While MSE can be used successfully in the OC-CF setting~\cite{koren2009matrix}, the Collaborative Denoising Autoencoder (CDAE) variant of AutoRec proposed a ranking loss alternative~\cite{wu2016collaborative}; we compare to CDAE in our experiments.

\subsection{Ranking Loss and Negative Sampling}
\label{sec:bpr}
Ranking loss explicitly considers the ranks of a user's historical observations, where observed interactions should have a higher ranking than unobserved interactions. In order to optimize rankings, a classic recommender system, Bayesian Pairwise Ranking (BPR)~\cite{rendle2009bpr}, 
proposes a generic approach called a triplet objective that maximizes the gap of prediction scores between the observed and unobserved interactions, which in its large-margin form is
\begin{equation}
\sum_{i,j} \max (r_{i,j}(\hat{r}_{i,j}-\hat{r}_{i,j'})+\epsilon, 0) \, .
\label{eq:bpr_loss}
\end{equation}
Here $j'$ is a randomly sampled item index and $\epsilon$ is the minimum margin of the prediction gap. The advantage of the triplet objective \cikm{over the MSE objective described in~\eqref{eq:autorec-loss}} is that it does not explicitly assume the unobserved interactions are negative examples.  Instead, it only makes an assumption on the priority of the observed over the unobserved interactions. 

Applying this sample-based triplet objective to Autoencoders was used by
%The sampling based training strategy for optimizing the triplet objective on an Autoencoder architecture is first introduced in 
Word2Vec~\cite{mikolov2013distributed}, which estimates the semantic meaning of words by exploiting word co-occurrence patterns.  Word2Vec leveraged Negative Sampling (NS) according to an attenuated popularity distribution.
With the success of word embeddings for natural language processing, researchers have also applied this training technique to many other application domains including recommender systems. For example, it is possible to achieve state-of-the-art performance on top-K recommendation tasks by simply borrowing the Word2Vec architecture without fundamental modifications~\cite{shin2014recommending}.

\subsection{Noise Contrastive Estimation}
\label{sec:nce}
Recent work (cf. \cite{bose-etal-2018-adversarial}, \cite{rao2016noise}) shows the triplet objective for ranking loss above is a simplified approximation of Noise Contrastive Estimation~(NCE)~\cite{gutmann2010noise}, where the predicted expectation of unobserved interactions are approximated with a single negative example. 

Formally, the Noise Contrastive Estimation objective in One-Class Collaborative Filtering (OC-CF) tasks can be defined as
\begin{equation}
\label{eq:nce-obj}
\begin{aligned}
% &{\arg\max}_{\hat{\mathbf{r}}_i} \sum_j \!r_{i,j} \!\left [\log \sigma(\hat{\mathbf{r}}_{i,j})\! +\!E_{p(j')}[\log\sigma( -\hat{\mathbf{r}}_{i,j'})]\right ],
&\argmax_{\tilde{\mathbf{r}}_i} \sum_j \!r_{i,j} \!\left [\log p(\tilde{r}_{i,j}=1)\! +\!E_{p(j')}[\log p(\tilde{r}_{i,j'}=0)]\right ],
\end{aligned}
\end{equation}
where we can simply assume the density
\begin{equation}
\label{eq:sigmoid-density}
    p(\tilde{r}_{i,j}=1) = \sigma(\tilde{r}_{i,j}) \quad \textrm{and}\quad p(\tilde{r}_{i,j'}=0) = 1 -\sigma(\tilde{r}_{i,j'})
\end{equation}
and the sampling probability $p(j')$ corresponds to the item popularity distribution with which we want to contrast users' preferences. 
In brief, the NCE objective optimizes for the prediction matrix $\tilde{R}$ that maximizes the gap of prediction scores between explicitly observed interactions and an expectation of unobserved interactions.

While optimizing NS or NCE objectives may seem attractive 
due to their previously established ``de-popularized'' embedding properties that improved overall recommendation performance~\cite{sigir19a}, this work also showed that
%in terms of enhancing personalization of the recommendation by optimizing the rank of items for each user, a recent work NCE-PLRec \cite{sigir19a} shows that
such approaches penalize popular items in a way that hurts general recommendation performance if used for the final task loss.
%that is estimated through information retrieval metrics. 
Thus, a combination of the NS or NCE objective for training embeddings along with an end task objective for recommendation is required to balance the need for high-quality embeddings with the need for good recommendations. 
However, it is not immediately clear how to combine two different objectives in AutoRec, given its end-to-end style of training.  We address this next by creating a two-headed AutoRec.

\section{Two-headed AutoRec Models}
While AutoRec is a strong baseline recommendation system, we will demonstrate empirically that it displays a bias to over-recommend popular items.  
Motivated by recent work that mitigated popularity bias through Negative Sampling (NS) and its closed-form noise contrastive estimation (NCE) generalization~\cite{sigir19a} by ``de-popularizing'' embeddings, we consider in this work whether such methods can be used to improve Autoencoder based methods.  

Specifically, we employ a two-headed AutoRec structure instead of standard AutoRec as seen in Figure \ref{fig:nce-autorec} (middle and right). The motivation of this structure naturally comes from the fact that while NS and NCE are good for training embeddings (where ``de-popularization'' of the embedding effectively leads to more nuanced embedding models across the popularity spectrum), neither NS or NCE is a good end loss for the final recommendation task. To resolve this, we make a separate head for each objective -- one to train the embedding via NS or NCE and the other to train for the recommendation task.  Next we introduce NS-AutoRec and NCE-AutoRec as two alternative approaches to combat popularity bias.

\subsection{NS-AutoRec}
\label{sec:nsautorec}
The Negative Sampling~(NS) enhanced Autoencoder based Recommender (NS-AutoRec) explicitly optimizes a personalized ranking objective through NS along with the Mean Square Error~(MSE) objective of AutoRec, as shown in Figure \ref{fig:nce-autorec} (middle). 
Since both of the heads in the proposed network architecture can produce valid recommendations, we distinguish them with different prediction variables to reduce notational confusion; we use $\hat{\mathbf{r}}_i$ to represent the prediction from the MSE objective, whereas we use $\tilde{\mathbf{r}}_i$ to represent the prediction from the NS objective.  What is critical in this architecture is that the two heads share a common user embedding.

Intuitively, we can jointly optimize the two objectives (one for each head) concurrently through a simple summation
\begin{equation}
    \argmin_{\theta,\vartheta,\psi} \mathcal{L}_{\theta,\vartheta}^{MSE} + \mathcal{L}_{\psi,\vartheta}^{NS},
\end{equation}
where $\vartheta$ denotes the encoder parameter set, and $\theta,\psi$ represent parameter sets of MSE and NS decoders, respectively.
However, comparing the two objectives, we note that the NS objective becomes a computational bottleneck during joint training due to 
(1)~sampling without replacement and (2)~the need to contrast observed and unobserved interactions~(sampled) for each user individually. 

In order to mitigate these drawbacks of inefficient sampling, we propose a few improvements.
%\subsubsection{Batch Negative Sampling} 
% TODO: as above, this is not clear.  What is "non-replacement sampling"?
First, sampling negative examples without replacement for each user at each training epoch is expensive.  Thus we alternatively sample the negative examples for all users 
at the same time \emph{with} replacement.  Second, instead of sampling from the uniform distribution, we sample negative examples from the item popularity distribution as commonly done in NS\footnote{\cikm{See  \cite{chen2018improving} for word embeddings and Section 3.6.2 of \cite{shlomo2018collaborative} for recommendation.}}
\begin{equation}
    p(j) = \frac{\norm{r_{:,j}}_1}{\sum_{j'}\norm{r_{:,j'}}_1},
\label{eq:nce-autorec_item_dist}
\end{equation}
which is computed via a simple normalization of item frequency in the training data.
%     and the parameter $k$ in Equation \ref{eq:nce-autorec_mult} indicates the number of times we sample for each user-item intersection. 

A consequence of this approach is that the entries of our sparse \emph{negative sample} matrix $S \in \mathbb{Z}^{m\times n}$ consist of integer counts instead of binary values.  By using this count representation of negatives, we dramatically reduce the number of times we need to sample during training. It is also worth noting that this efficient and compact representation of negatives also necessitates a custom specification of the contrastive objective used by NS explained further below.

The usual NS objective is a triplet loss as shown in Equation~\ref{eq:bpr_loss} that maximizes the gap between positive observations and negatives sampled from
the item popularity distribution $p(\mathcal{J})$.
Because in our setting, there is not a single negative sample for each explicitly observed positive example, we propose an alternative aggregated objective function that explicitly trains on all positive observations and negative samples at one time. 
% This is NS, not NCE (which is in the next subsection)... why is NCE coming up here... it's quite jarring for the reader.
%the multiple training epochs into one where mini-batch of NS is optimized instead of stochastic optimization. 
Concretely, we define the objective function for the NS head of NS-AutoRec as follows:
\begin{equation}
\begin{split}
&
%\argmax_{\psi,\vartheta} 
\mathcal{L}_{\psi,\vartheta}^{NS} =
\sum_i \left [\mathbf{r}_i^T (f_\psi \circ f_\vartheta(\mathbf{r}_i)) - \frac{\norm{\mathbf{r}_i}_1}{\norm{\mathbf{s}_i}_1} \mathbf{s}_i^T  (f_\psi \circ f_\vartheta(\mathbf{r}_i)) \right ].
\end{split}
\label{eq:nce-autorec_nce_obj_modified}
\end{equation}
While it appears complex at first, the intuition for this form is quite simple.  The first term effectively counts (i.e., via an inner product with a 0-1 sparse vector) the number of positive item observations in the data that the Autoencoder is able to reconstruct.  However, in the NS approach we also want to penalize the Autoencoder for predicting high scores for negative samples, which can be done by a second negated inner product, this time with $\mathbf{s}_i^T$ which is simply the vector of negative sample counts for each item for user $i$.  The term $\frac{\norm{\mathbf{r}_i}_1}{\norm{\mathbf{s}_i}_1}$
simply balances the contribution of the two terms to contribute equally to the objective even if the number of positive examples and negative samples does not match.  In summary, Equation~\eqref{eq:nce-autorec_nce_obj_modified} gives us a compact and efficiently trainable NS-AutoRec. 

For the objective of the second head of NS-AutoRec, we simply minimize Mean Squared Error (MSE):
\begin{equation}
\begin{split}
&
%\argmin_{\theta,\vartheta} 
\mathcal{L}_{\theta,\vartheta}^{MSE} =
\sum_i \norm{\mathbf{r}_i - (f_\theta \circ f_\vartheta(\mathbf{r}_i)) }_2,
\end{split}
\label{eq:nce-autorec_mle_obj}
\end{equation}
where the variable set $\theta$ contains the parameters of the MSE decoding network. Note the encoding network $f_\vartheta$ is intentionally shared for both decoder loss objectives $\mathcal{L}_{\psi,\vartheta}^{NS}$ and $\mathcal{L}_{\theta,\vartheta}^{MSE}$.

We will see empirically that NS-AutoRec does produce improved embeddings for AutoRec that reduce popularity bias.  However, it turns out that NS can be generalized in a Noise Contrastive Estimation (NCE) framework that yields both faster training times as well as improved performance.  We cover this second variant of two-headed AutoRec next.

\subsection{NCE-AutoRec}
\label{sec:nce-autorec_nce_description}
In this section, we propose an alternative two-headed Autoencoder model called NCE-AutoRec, as shown in Figure \ref{fig:nce-autorec}~(right). 

Like NS-AutoRec, NCE-AutoRec also jointly minimizes the loss based on the sum of objectives for each head: 
\begin{equation}
    \argmin_{\theta,\vartheta,\phi} \mathcal{L}_{\theta,\vartheta}^{MSE} + \mathcal{L}_{\phi,\vartheta}^{NCE}.
\end{equation}
As before, we use $\hat{\mathbf{r}}_i$ to represent the prediction from the MSE objective, whereas we use $\tilde{\mathbf{r}}_i$ to represent the prediction from the NCE objective. 

We start our discussion with the NCE objective head. Recall the NCE objective in its original form
\begin{equation}
\begin{split}
&\argmax_{\phi, \vartheta} \sum_i \sum_j r_{i,j} \left [\log\sigma(\tilde{r}_{i,j}) + E_{p(j')}[\log\sigma(-\tilde{r}_{i,j'})]\right ],
\end{split}
\label{eq:nce-autorec_original}
\end{equation}
where the prediction of NCE head is
\begin{equation}
    \tilde{\mathbf{r}}_i = f_\phi \circ f_\vartheta(\mathbf{r}_i)
\end{equation}
and scalar $\tilde{r}_{i,j}$ represents the $j$th entry of the vector $\tilde{\mathbf{r}}_i$. Similar to  NS-AutoRec described previously, the item probability $p(j')$ is a re-scaled empirical item popularity as described in Equation \ref{eq:nce-autorec_item_dist}. 

We now reformulate Equation~\ref{eq:nce-autorec_original} as a constrained optimization of the form:
\begin{equation}
\begin{split}
\argmax_{\phi, \vartheta} &\underbrace{\sum_i \sum_j r_{i,j} \left [\log\sigma(\tilde{r}_{i,j}) + E_{p(j')}[\log\sigma(-\tilde{r}_{i,j'})]\right ]}_{\textcircled{1}: \ell}\\
&\underbrace{-\sum_i\lambda_i\norm{\tilde{\mathbf{r}}_i - f_\phi \circ f_\vartheta(\mathbf{r}_i)}_2}_{\textcircled{2}},
\end{split}
\label{eq:nce-autorec_optimization}
\end{equation}
where $\lambda_i$ is a Lagrange multiplier for a term encouraging equality between $\tilde{\mathbf{r}}_i$ and $f_\phi \circ f_\vartheta(\mathbf{r}_i)$. This reformulation allows us to optimize a lower-bound of the objective function by maximizing the two components $\textcircled{1}$ and $\textcircled{2}$ separately.

Now we can obtain the optimal solution of the objective $\textcircled{1}$ by computing the derivative of the objective with respect to the prediction $\tilde{r}_{i,j}$. Formally, the derivative is 
\begin{equation}
\begin{split}
\frac{\partial \ell }{\partial \tilde{r}_{i,j} } &= \sigma(-\tilde{r}_{i, j}) - \frac{\norm{r_{:,j}}_1}{\sum_{j'}\norm{r_{:,j'}}_1} \sigma(\tilde{r}_{i, j}).
\end{split}
\end{equation}
and has the closed-form optimal solution for an observed interaction
% The corresponding optimal solution $r_{i,j}^*$  for an observed interaction is
\begin{equation}
r_{i,j}^* = \log \frac{\sum_{j'}\norm{r_{:,j'}}_1}{\norm{r_{:,j}}_1} \qquad \forall r_{i,j} = 1,
\label{eq:r_star}
\end{equation}
and is simply zero for an unobserved interaction:
\begin{equation}
r_{i,j}^*= 0 \qquad \forall r_{i,j} = 0.
\label{eq:r_star_neg}
\end{equation}

\label{sec:beta}
As one variation, we remark that weighting the denominator of Equation~\eqref{eq:r_star} with a hyper-parameter $\beta$ allows us to adjust for inaccuracies in popularity estimates and lower bounding it at zero ensures that positive examples always outweigh negative feedback.
Both prove useful in our experimental evaluation:
\begin{equation}
r_{i,j}^* = max(\log \sum_{j'} \norm{r_{:,j'}}_1 - \beta\log \norm{r_{:,j}}_1, 0) \qquad \forall r_{i,j} = 1.
\label{eq:nce_alter_solution}
\end{equation}

With the analytical solution $R^*$ of the NCE objective, we, then, aim to maximize the objective component $\textcircled{2}$ to reduce the gap between the Autoencoder prediction and the analytical solution $R^*$. Specifically, we train a regression model with
\begin{equation}
\begin{split}
\mathcal{L}_{\phi,\vartheta}^{NCE} = \sum_i \norm{\mathbf{r}_i^* - (f_\phi \circ f_\vartheta(\mathbf{r}_i)) }_2,
\end{split}
\label{eq:nce-autorec_nce_obj_closed}
\end{equation}
which serves as the NCE head objective in NCE-AutoRec.

The MSE objective of NCE-AutoRec is identical to the one in NS-AutoRec, as shown in Equation~\eqref{eq:nce-autorec_mle_obj}.

For both NS-AutoRec and NCE-AutoRec, optimizing two objectives simultaneously is difficult because of the involvement of many hyper-parameters. 
We will next discuss more details of optimization approaches for these two-headed models in Section \ref{sec:opt}.

\subsection{Optimization Methodology for NS-AutoRec and NCE-AutoRec} \label{sec:opt}
We now describe four possible ways to optimize parameters of the two heads of NS-AutoRec and NCE-AutoRec as follows:

\begin{itemize}
    \item \textbf{Joint}: We optimize all parameters through both objectives (heads and losses) jointly. Intuitively, we can jointly optimize the sum of NCE~(or NS) and MSE objectives with stochastic gradient descent. 
    \item \textbf{Alternating}: We alternatively optimize each of the objectives. 
    Here, we first optimize parameters w.r.t.\ the NCE (or~NS) objective and then optimize w.r.t.\ the MSE objective for each batch of input.
    \item \textbf{Limited Fine-tune}: We optimize the NCE~(or NS) objective until it is fully converged. We then optimize an MSE objective that blocks backpropagation to the encoder network by freezing the encoder weights.
    By doing this, we avoid providing misleading embedding information to the shared embedding layer (as the performance results indicate later).
    \item \textbf{Full Fine-tune}: We optimize the NCE~(or NS) objective until it is fully converged. We then optimize the MSE objective without blocking backpropagation. This allows the MSE objective to fine-tune the embeddings learned through the NCE~(or NS) objective.
\end{itemize}

We treat the choice of \textbf{Joint}, \textbf{Alternating}, \textbf{Limited Fine-tune} and \textbf{Full Fine-tune} optimization methodologies as a training hyperparameter for NS-AutoRec and NCE-AutoRec. As empirical results later verify in Figure \ref{fig:nsautorec_nceautorec_ndcg}, \textbf{Limited Fine-tune} consistently works the best for NCE-AutoRec whereas \textbf{Joint} is very competitive in larger datasets. \textbf{Full Fine-tune} and \textbf{Alternating} give promising results for NS-AutoRec in specific datasets.

\section{Experiments}
In this section, we evaluate the performance of NS-AutoRec and NCE-AutoRec by comparing them with several classic and state-of-the-art baselines. We begin by describing our datasets and experimental settings, and then analyze the results in order to address the following research questions \cikm{with a particular emphasis on assessing the level of personalization (i.e., avoidance of a strong popularity bias) in NS-AutoRec and NCE-AutoRec}:

\begin{itemize}
    \item \textbf{RQ1}: How do NS-AutoRec and NCE-AutoRec perform against state-of-the-art methods?
    \item \textbf{RQ2}: How do different optimization settings, number of negative samples and popularity sensitivity affect the performance of NS-AutoRec and/or NCE-AutoRec?
    \item \textbf{RQ3}: How do NS-AutoRec and NCE-AutoRec perform on personalized recommendation and a cold-start user study?
    \item \textbf{RQ4}: How efficient are NS-AutoRec and NCE-AutoRec in terms of training time under different optimization settings?
\end{itemize}

\iffalse
We primarily show that
\begin{enumerate}
    \item NCE-AutoRec outperforms NS-AutoRec in all evaluation metrics on three real-world large scale datasets (Table \ref{table:ml-20m}, \ref{table:netflix} and \ref{table:yahoo}).
    \item Both NS-AutoRec and NCE-AutoRec allow us to provide more personalized recommendation compared to other Autoencoder based algorithms and naive popularity based recommendation as illustrated in Figure \ref{fig:popularity}.
    \item While NS-AutoRec improves results compared to AutoRec and achieves comparable results with NCE-AutoRec in some cases, it is very inefficient in applying to large datasets as Negative Sampling takes up a majority of the runtime as shown in Figure \ref{fig:runtime}.
\end{enumerate}
\fi

\subsection{Datasets}

\begin{table}
  \caption{Summary statistics of datasets used in evaluation}
  \label{tab:datasets}
  \resizebox{\columnwidth}{!}{%
  \begin{tabular}{ccccc}
    \toprule
    Dataset & $m$ & $n$ & |$r_{i, j} > \eta$| & Sparsity \\
    \midrule
    Goodbooks & 53,418 & 10,000 & 5,493,027 & $1.03\times10^{-2}$ \\
    MovieLens-20M & 138,362 & 22,884 & 12,195,566 & $3.85\times10^{-3}$ \\
    Netflix & 478,615 & 17,769 & 56,919,192 & $6.69\times10^{-3}$ \\
    Yahoo & 1,876,280 & 44,865 & 48,817,552 & $5.80\times10^{-4}$\\
    \bottomrule
  \end{tabular}
  }
\vspace{-5mm}
\end{table}

We use four large-scale benchmark recommendation datasets in our experiments: Goodbooks \cite{goodbooks2017}, MovieLens-20M \cite{Harper:2015:MDH:2866565.2827872}, Netflix \cite{bennett2007netflix} and Yahoo \cite{Dror:2011:YMD:3000375.3000377}. They are publicly accessible and vary in domain, size and sparsity. The summary statistics for the four datasets are in Table \ref{tab:datasets}.

\begin{itemize}
    \item \textbf{Goodbooks}: Goodbooks dataset, the first of its kind, is a book recommendation dataset.
    \item \textbf{MovieLens-20M}: MovieLens-20M dataset is a stable and widely-used dataset for movie recommendation. 
    \item \textbf{Netflix}: Netflix dataset comes from the well-known Netflix Prize dataset for movie recommendation.
    \item \textbf{Yahoo}: Yahoo dataset represents a snapshot of music preferences of Yahoo Music Community.
\end{itemize}

\cikm{We follow a similar data split as seen in \cite{sigir19a}. For each user of each dataset, we use 50\% interactions as training set, 20\% as validation set and 30\% as test set according to timestamps.} We split the Goodbooks and Yahoo dataset randomly due to lack of timestamps. We set a threshold $\eta$ to binarize the ratings in each dataset such that ratings greater or equal than $\eta$ become 1 and otherwise 0. The threshold is 80 for Yahoo dataset and 3 for the rest.

\subsection{Experimental Settings}

\subsubsection{Baselines}

We compare our proposed NS-AutoRec and NCE-AutoRec with the following \cikm{recommender systems that can scale up to industry-level applications with millions of users and items}:

\begin{itemize}
  \item \cikm{\textbf{BPR} \cite{rendle2009bpr}: Bayesian Pairwise Ranking that optimizes the pairwise ranking objective discussed in Section~\ref{sec:bpr}.}
  \item \textbf{CDAE} \cite{wu2016collaborative}: Collaborative Denoising Autoencoder that is optimized for implicit feedback recommendation scenarios.
  \item \textbf{CML} \cite{hsieh2017collaborative}: Collaborative Metric Learning. A state-of-the-art metric learning based method.
  \item \textbf{NCE-PLRec} \cite{sigir19a}: PLRec \cite{Sedhain:2016:PLM:3061053.3061158} variant where item representation is learned from NCE.
  \item \textbf{PureSVD} \cite{cremonesi2010performance}: Similarity based recommendation method by SVD decomposition of $R$.
  \item \textbf{VAE-CF} \cite{liang2018variational}: Variational Autoencoder for CF, which is a state-of-the-art deep learning based method.
  \item \textbf{WRMF} \cite{koren2009matrix}: Weighted Regularized Matrix Factorization.
  \item \textbf{AutoRec} \cite{Sedhain:2015:AAM:2740908.2742726}: One-headed Autoencoder based method that contains an encoding layer and decoding layer with ReLU activation function optimized with MSE objective. This is one ablation of NS-AutoRec and NCE-AutoRec.
  \item \textbf{OHNS-AutoRec}: One-headed Autoencoder based method that is the same as AutoRec except optimized with Negative Sampling objective. This is another ablation of NS-AutoRec.
  \item \textbf{NS-AutoRec}: Two-headed Autoencoder based method we proposed that optimizes one head with Negative Sampling objective, and the other head with MSE objective.
  \item \textbf{NCE-AutoRec}: Two-headed Autoencoder based method we proposed that optimizes one head with NCE objective through closed-form solution, and the other head with MSE objective.
\end{itemize}

\subsubsection{Hyper-parameter Settings}
We implement all methods in TensorFlow\footnote{https://www.tensorflow.org} and tune all the hyper-parameters by evaluating methods on the validation set with grid search. We tune the following parameters:
\begin{itemize}
    \item $r$: (latent dimension) from \{50, 100, 200, 500\} for all methods.
    \item $\lambda$: (regularization strength) from \{1e-7, 1e-6 · · · 1e4\} for all methods.
    \item $\alpha$: (loss weighing) from \{-0.5, -0.4 · · · -0.1\} $\cup$ \{0, 0.1, 1, 10, 100\} for NCE-PLRec and WRMF.
    \item $\rho$: (corruption) from \{0.1, 0.2 · · · 1\} for CDAE and VAE-CF.
    \item $\beta$: (popularity pensitivity) from \{0.7, 0.8 · · · 1.3\} for NCE-PLRec and NCE-AutoRec.
    \item $N$: (number of negative samples) from \{5, 50, 500, 5000, 50000, 500000\} for OHNS-AutoRec and NS-AutoRec.
    \item \textit{mode}: (optimization methodology) from \{\textbf{Joint}, \textbf{Alternating}, \textbf{Limited fine-tune}, \textbf{Full Fine-tune}\} for NS-AutoRec and NCE-AutoRec.
\end{itemize}

\begin{figure*}[t]%
\includegraphics[width=0.6\linewidth]{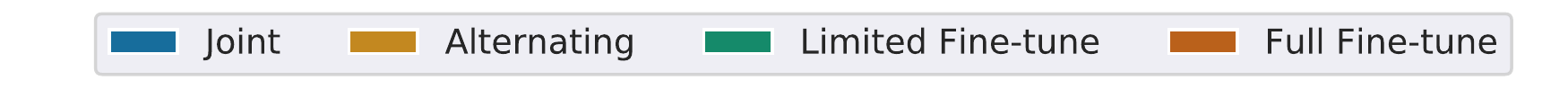}
\vspace{-2mm}
\begin{subfigure}{0.5\linewidth}
\includegraphics[width=\linewidth]{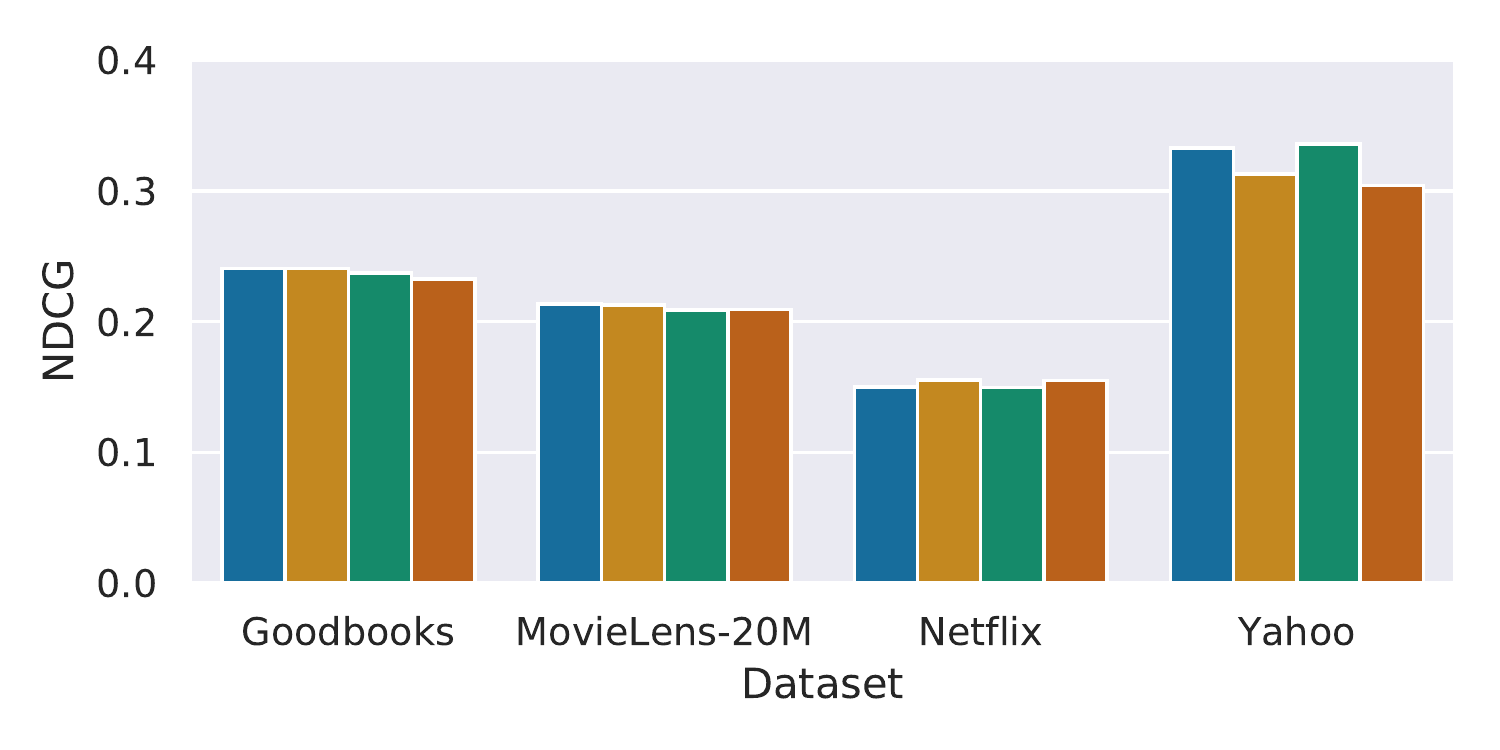}
\vspace{-2\baselineskip}
\caption{NS-AutoRec}
\label{fig:nsautorec_nceautorec_ndcg_a}%
\end{subfigure}\hfill%
\begin{subfigure}{0.5\linewidth}
\includegraphics[width=\linewidth]{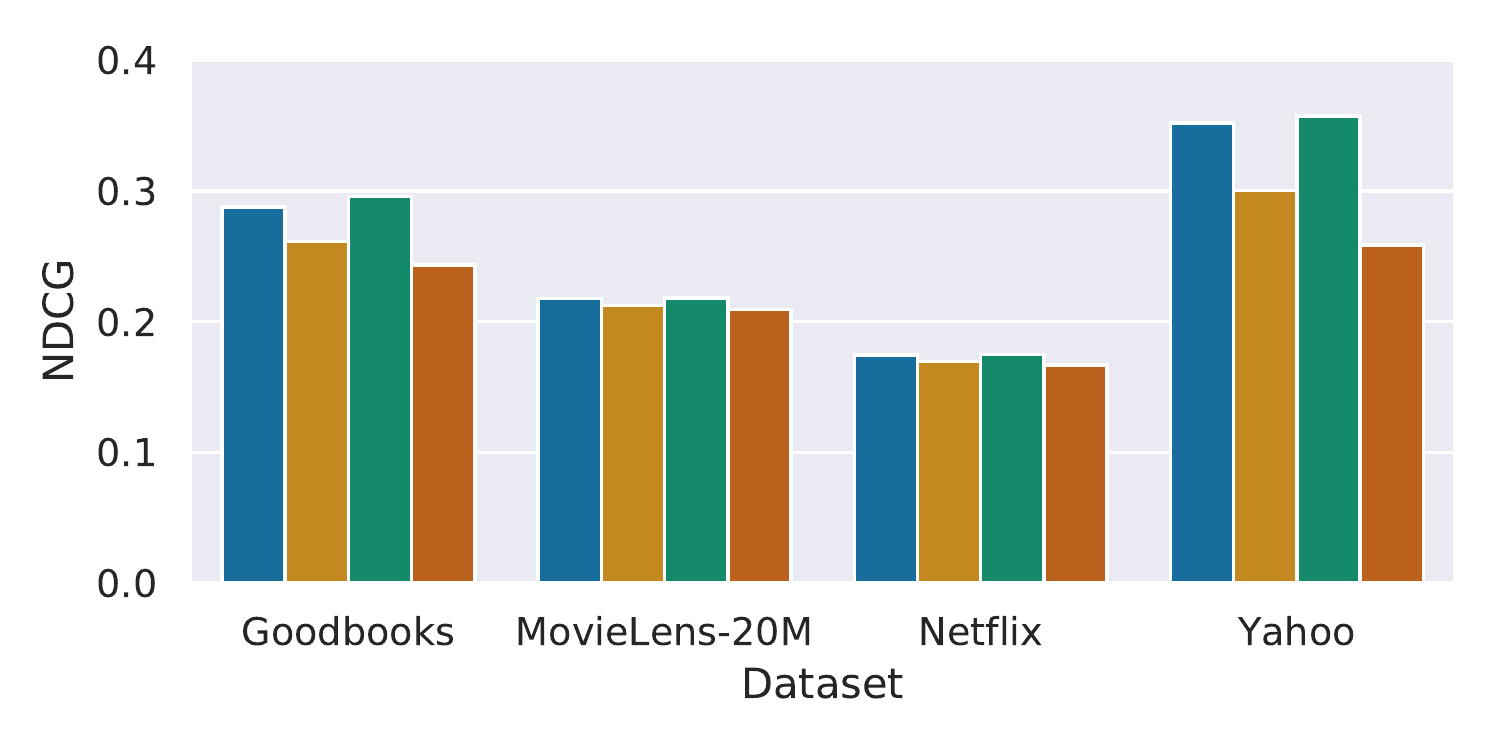}
\vspace{-2\baselineskip}
\caption{NCE-AutoRec}
\label{fig:nsautorec_nceautorec_ndcg_b}%
\end{subfigure}\hfill%
\caption{NDCG comparison for NS-AutoRec and NCE-AutoRec with different optimization settings}
\label{fig:nsautorec_nceautorec_ndcg}
\end{figure*}

\begin{figure*}[t]%
\begin{subfigure}{\linewidth}
\includegraphics[width=\linewidth]{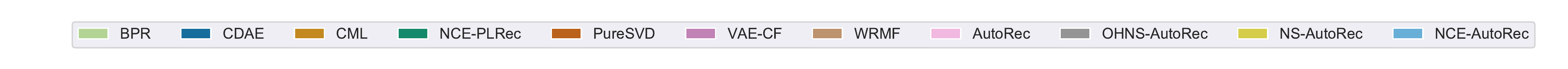}%
\vspace{-2mm}
\end{subfigure}%

\begin{subfigure}{0.5\linewidth}
\includegraphics[width=\linewidth]{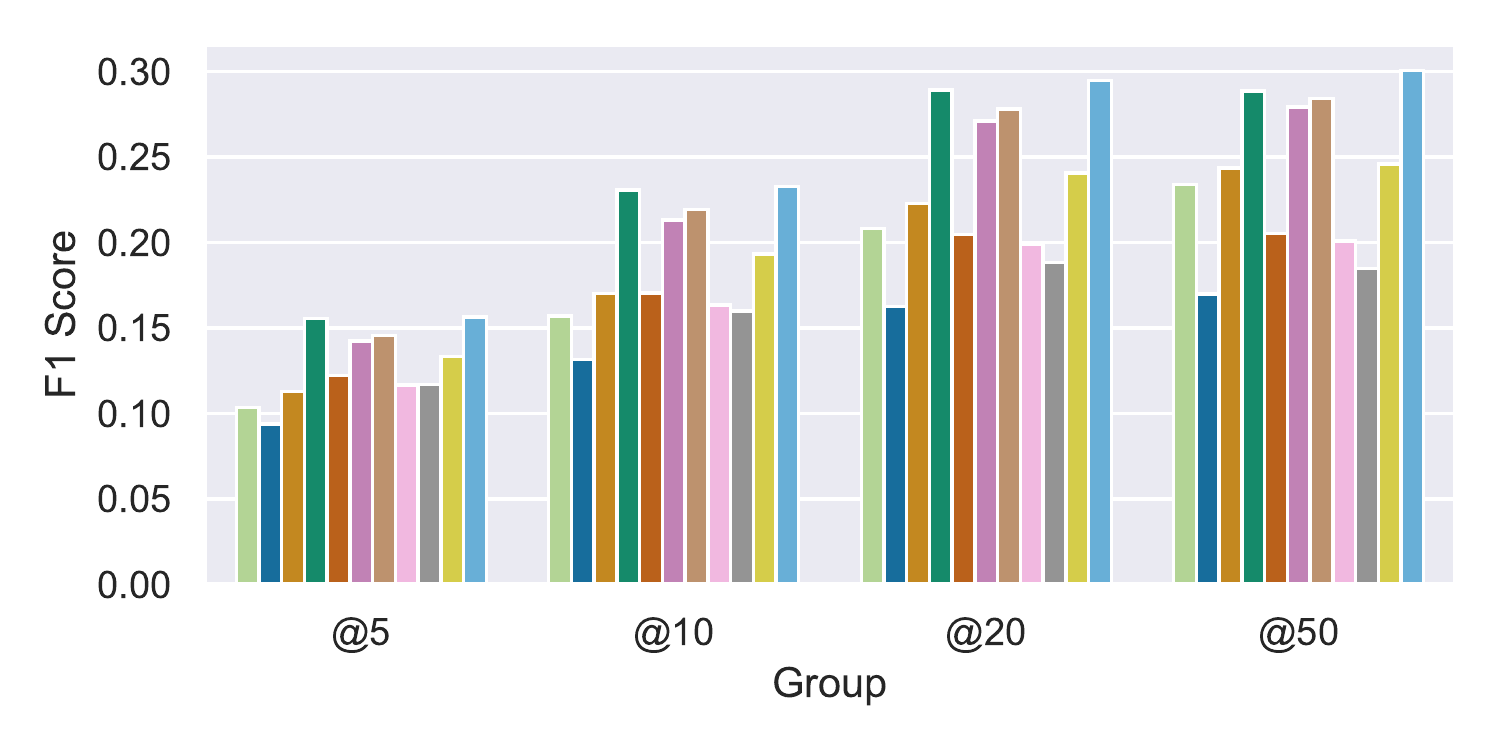}%
\vspace{-4mm}
\caption{Goodbooks}%
\end{subfigure}\hfill%
\begin{subfigure}{0.5\linewidth}
\includegraphics[width=\linewidth]{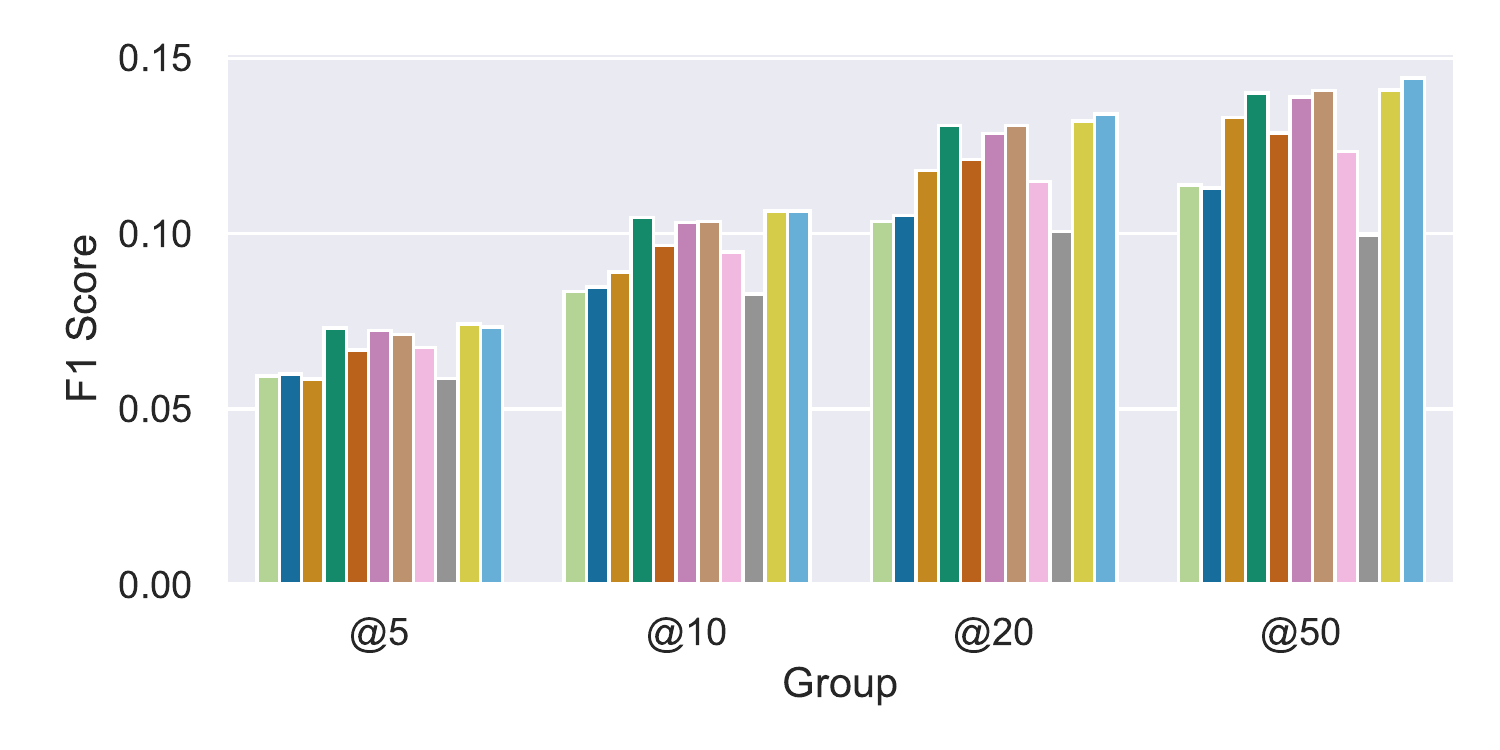}%
\vspace{-4mm}
\caption{MovieLens-20M}%
\end{subfigure}\hfill%
\begin{subfigure}{0.5\linewidth}
\includegraphics[width=\linewidth]{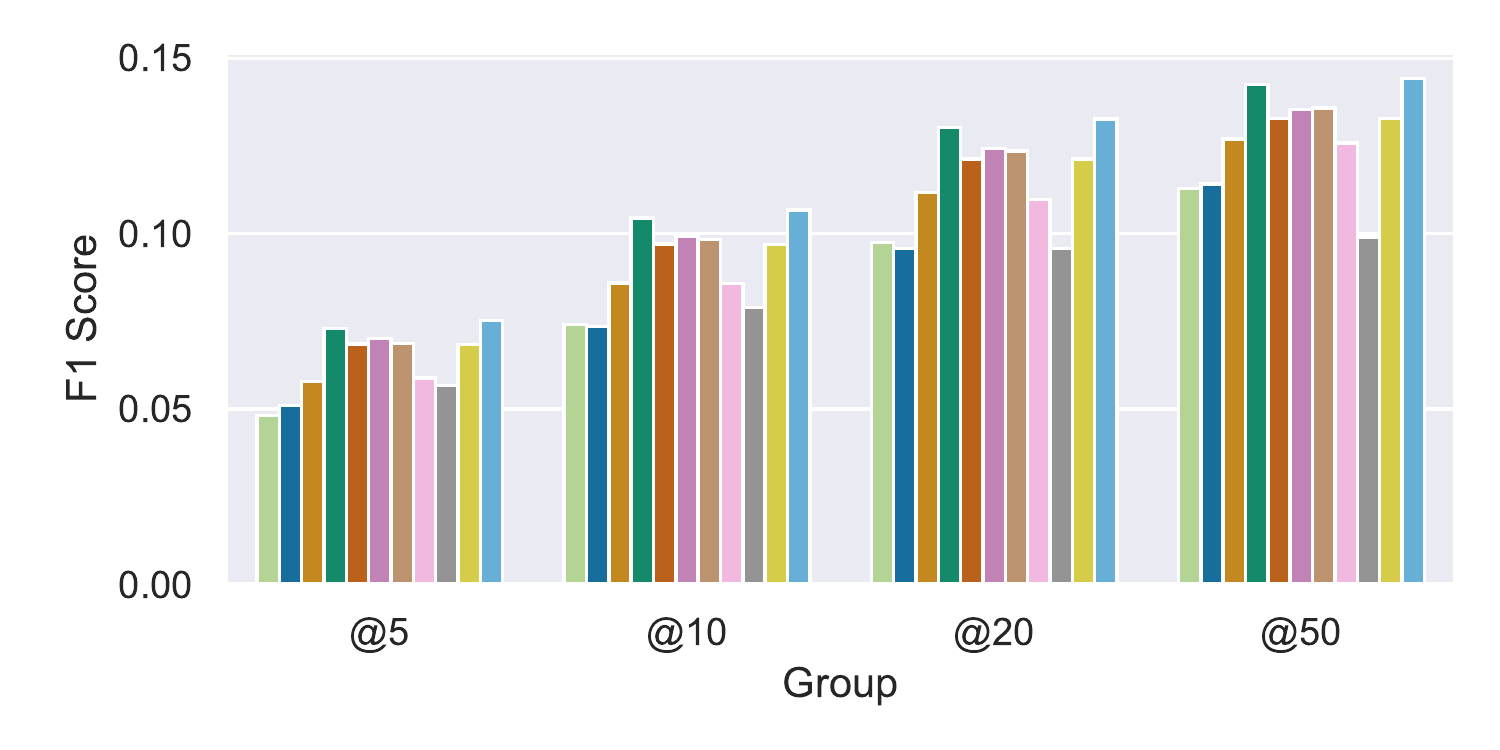}%
\vspace{-4mm}
\caption{Netflix}%
\end{subfigure}\hfill%
\begin{subfigure}{0.5\linewidth}
\includegraphics[width=\linewidth]{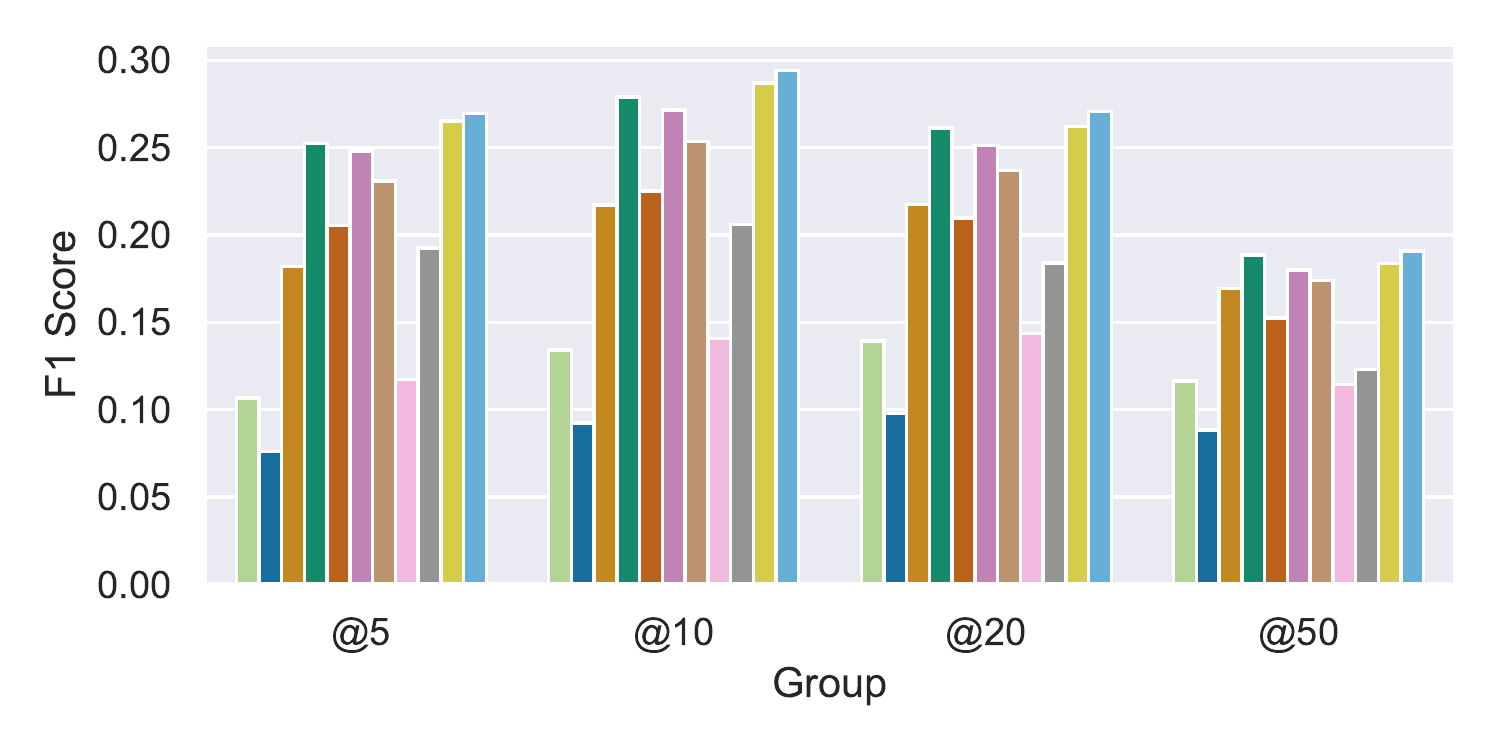}%
\vspace{-4mm}
\caption{Yahoo}%
\end{subfigure}\hfill%
\centering
\caption{F1 score bar plots for baseline recommendation methods for $K \in [5, 10, 20, 50]$. Taller bar indicates higher F1 score and NCE-AutoRec outperforms other baselines across the four datasets}
\label{fig:F1}
\end{figure*}

\begin{figure*}[t]%
\includegraphics[width=0.6\linewidth]{mode_legend.pdf}
\vspace{-2mm}
\begin{subfigure}{0.5\linewidth}
\includegraphics[width=\linewidth]{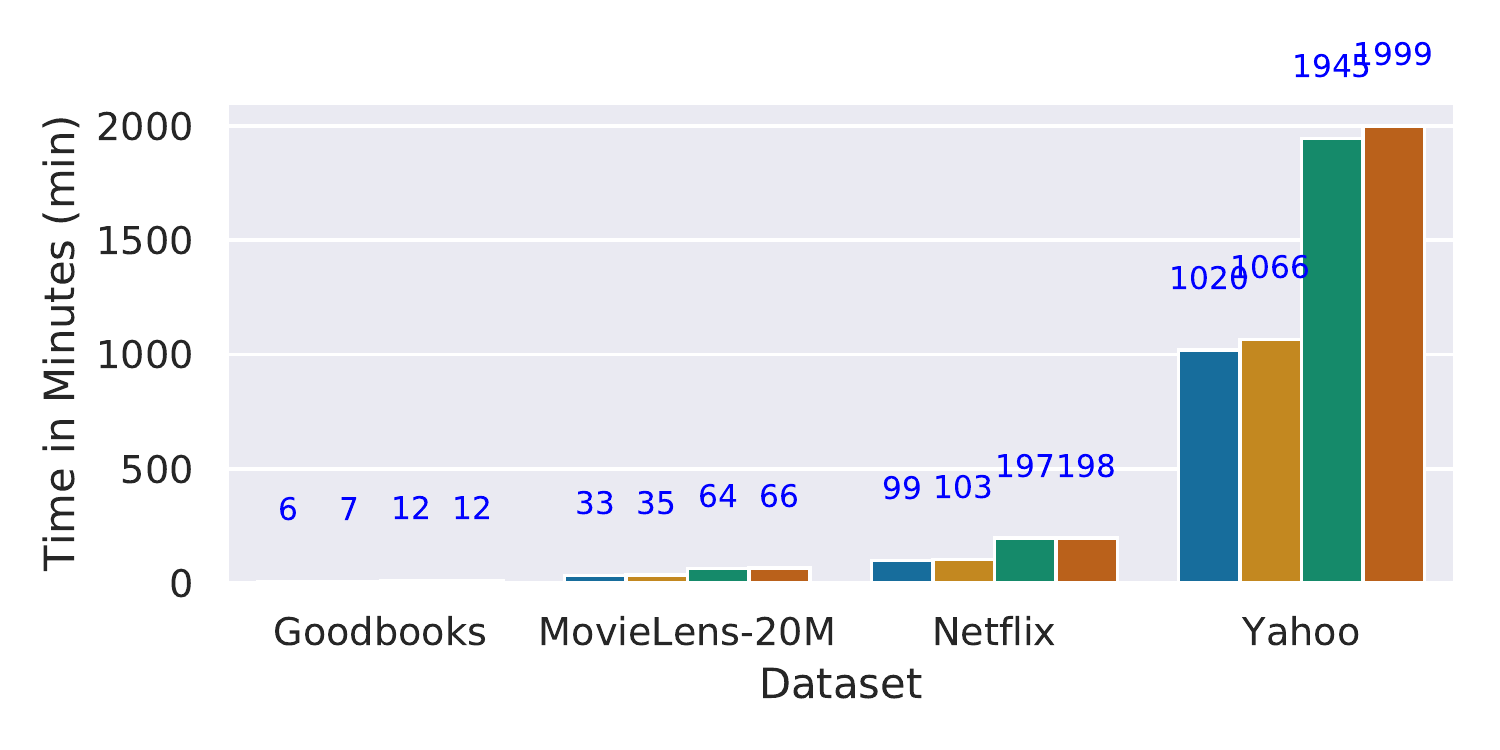}
\vspace{-2\baselineskip}
\caption{NS-AutoRec}
\label{fig:nsautorec_nceautorec_time_a}
\end{subfigure}\hfill%
\begin{subfigure}{0.5\linewidth}
\includegraphics[width=\linewidth]{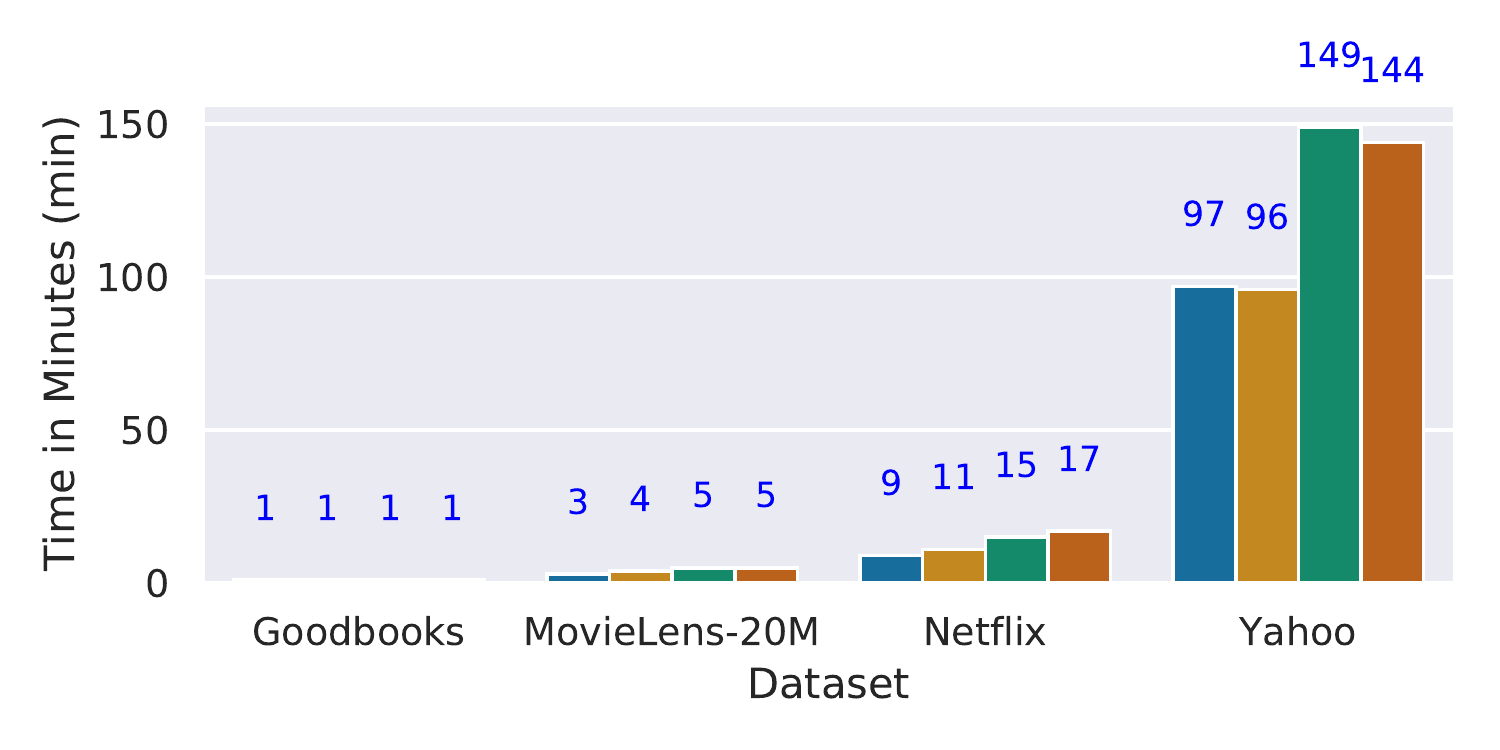}
\vspace{-2\baselineskip}
\caption{NCE-AutoRec}
\label{fig:nsautorec_nceautorec_time_b}%
\end{subfigure}\hfill%
\caption{Convergence time comparison for NS-AutoRec and NCE-AutoRec with different optimization settings}
\label{fig:nsautorec_nceautorec_time}
\end{figure*}

\begin{figure*}[t]%
\begin{subfigure}{0.5\linewidth}
\includegraphics[width=\linewidth]{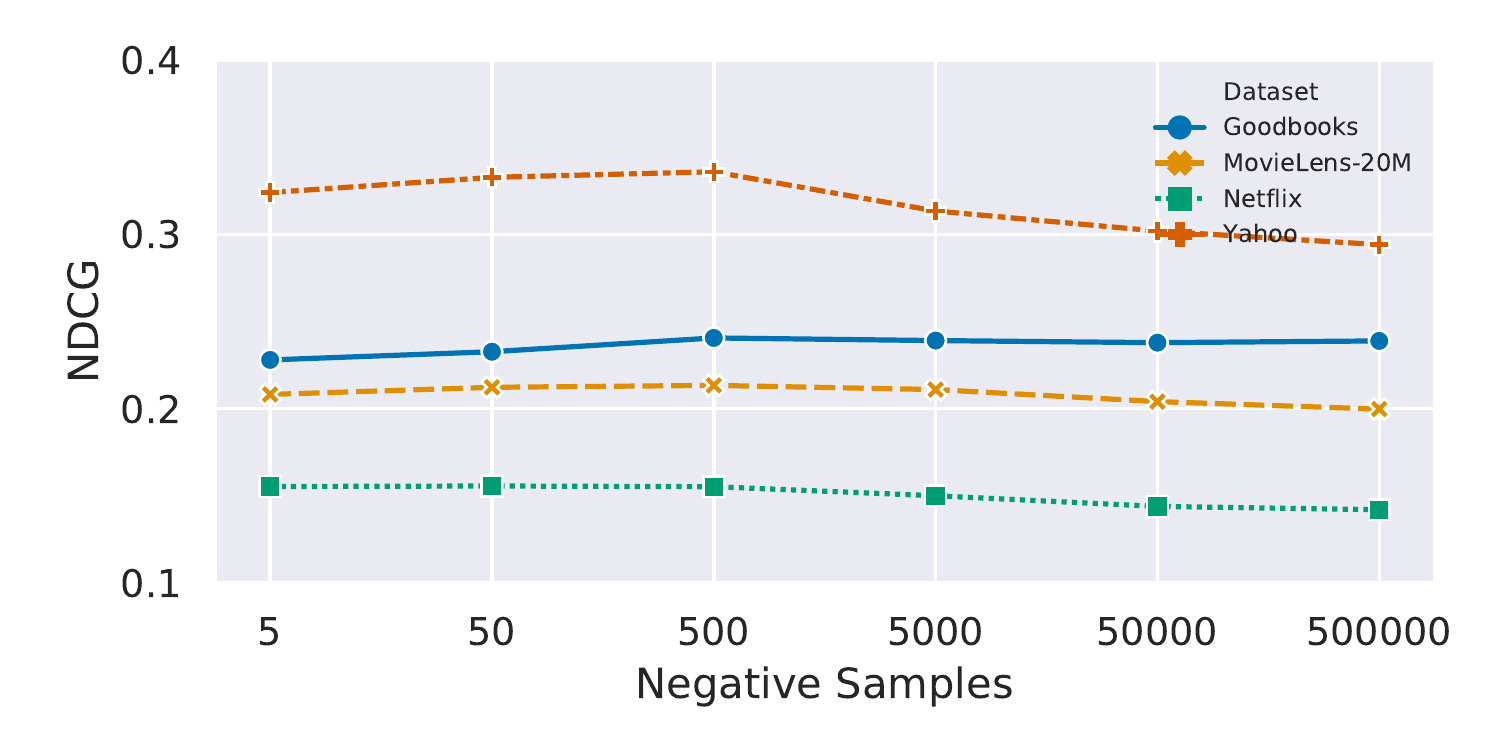}%
\caption{NS-AutoRec}
\label{fig:n_beta_tuning_a}%
\end{subfigure}\hfill%
\begin{subfigure}{0.5\linewidth}
\includegraphics[width=\linewidth]{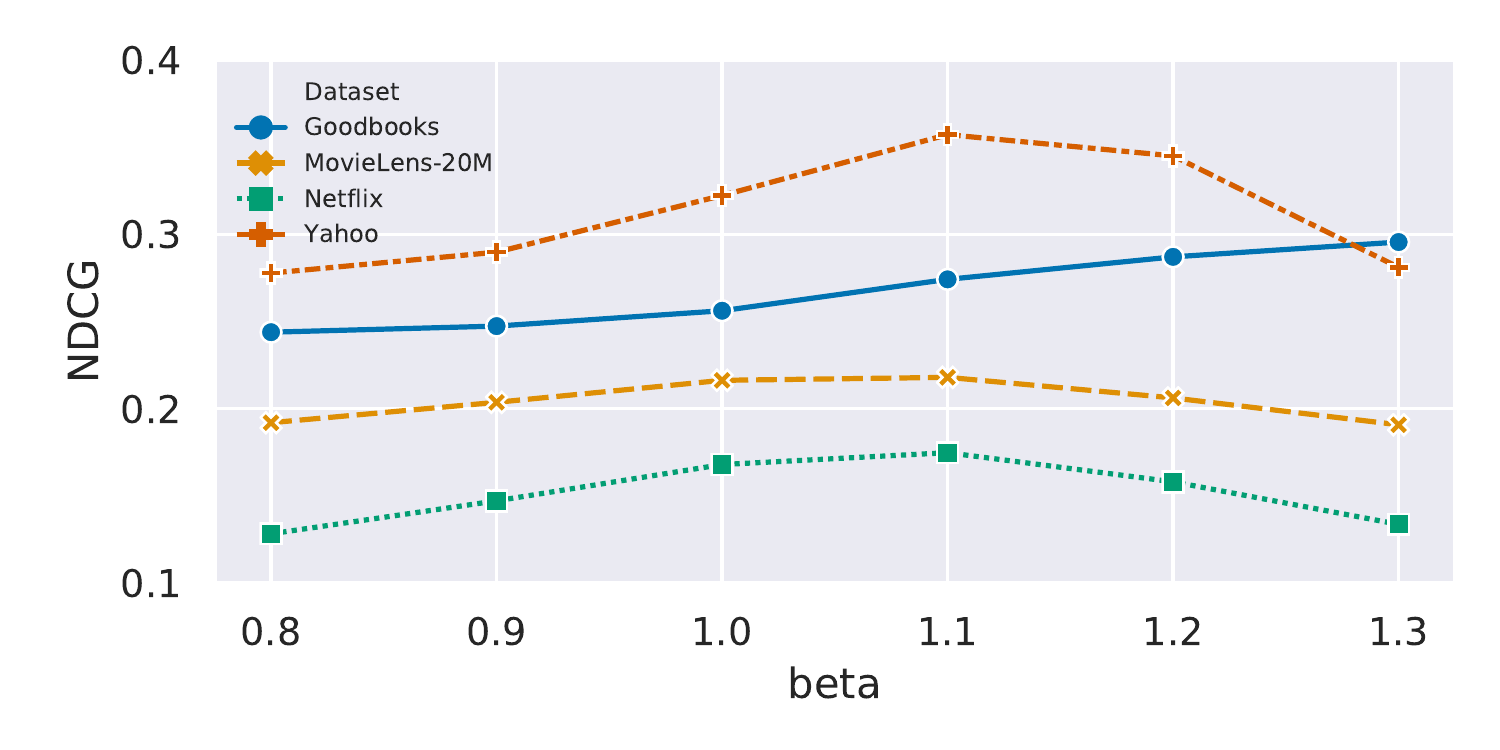}
\caption{NCE-AutoRec}
\label{fig:n_beta_tuning_b}%
\end{subfigure}\hfill%
\caption{Effect for NDCG of tuning $N$ on NS-AutoRec and $\beta$ on NCE-AutoRec}
\label{fig:n_beta_tuning}
\end{figure*}

\subsection{General Recommendation Performance}
To answer RQ1, we evaluate the performance of all methods using six standard evaluation metrics: R-Precision, NDCG, MAP@K, Precison@K, Recall@K and F1-score@K. The evaluation results of all methods on the test set are summarized in Tables \ref{table:goodbooks}, \ref{table:ml-20m}, \ref{table:netflix}, \ref{table:yahoo} \cikm{and reported with 95\% confidence intervals.} The F1-score of all methods is shown in Figure \ref{fig:F1} with $K \in [5, 10, 20, 50]$. The results demonstrate that NCE-AutoRec consistently outperforms all other baseline methods across the four domains.

While Negative Sampling has proven useful for OC-CF, it is apparent that Negative Sampling itself is not ideal for the end task as OHNS-AutoRec demonstrates a relatively poor performance across all four domains. It can be seen that by combining the Negative Sampling and MSE objectives together as a two-headed Autoencoder, NS-AutoRec outperforms two of its ablations: AutoRec and OHNS-AutoRec (by a large margin) and is strongly competitive compared to the other baselines. This verifies the effectiveness of the two-headed structure we proposed. Similarly, NCE-AutoRec also exhibits a large performance gain from AutoRec and OHNS-AutoRec and outperforms NS-AutoRec consistently.

\subsection{Impact of Hyper-parameter Tuning}
In this section, we study RQ2 by analyzing how different optimization settings (\textit{mode}), number of negative samples ($N$) and popularity sensitivity ($\beta$) introduced in Section \ref{sec:opt}, \ref{sec:nsautorec} and \ref{sec:beta} respectively affect the performance of NS-AutoRec and/or NCE-AutoRec. We use NDCG for the performance measure since other performance measures behave similarly.

\subsubsection{Optimization Settings} 
We analyze the performance of NS-AutoRec and NCE-AutoRec with different optimization settings across the four datasets. Specifically, we plot NDCG for NS-AutoRec and NCE-AutoRec with a different optimization $mode$ in Figure \ref{fig:nsautorec_nceautorec_ndcg}. Generally speaking, the best optimization $mode$ for NS-AutoRec depends on the dataset, suggesting the importance of including this hyper-parameter to achieve the best performance. For NCE-AutoRec, as Figure \ref{fig:nsautorec_nceautorec_ndcg_b} illustrates, \textbf{Joint} and \textbf{Limited Fine-tune} outperform other settings consistently. \textbf{Joint} performs noticeably worse than \textbf{Limited Fine-tune} in Goodbooks and slightly worse than \textbf{Limited Fine-tune} in larger datasets. This might suggest the fact that obtaining a better item embedding first as in \textbf{Limited Fine-tune} is more crucial in smaller datasets whereas in larger datasets jointly training two objectives does not hurt the item embedding too much and consequently the performance remains competitive.

\subsubsection{Number of Negative Samples}
To study the effect of $N$ on the performance of NS-AutoRec, we plot NDCG against $N$ on the datasets as shown in Figure \ref{fig:n_beta_tuning_a}. We observe that, across all datasets, increasing $N$ initially boosts the performance until $N=500$ and afterwards hurts NDCG as $N$ further increases to 500,000.  Since the positive and negative components of the optimization objective are balanced, we conjecture that a large $N$ simply spreads learning too thinly over a diverse set of negative samples, effectively adding a high level of noise to the learning process.

\subsubsection{Popularity Sensitivity}
The effect of tuning $\beta$ for NCE-AutoRec is shown in Figure \ref{fig:n_beta_tuning_b}. We observe a large improvement in performance for the least sparse dataset (Goodbooks) and the most sparse (Yahoo) dataset, whereas the performance boost is relatively small in the other two datasets.  This demonstrates that tuning the popularity sensitivity hyperparameter can be important for optimal performance in diverse dataset settings.

\subsection{Personalization}
\begin{figure}[t]
  \centering
  \includegraphics[width=0.9\linewidth]{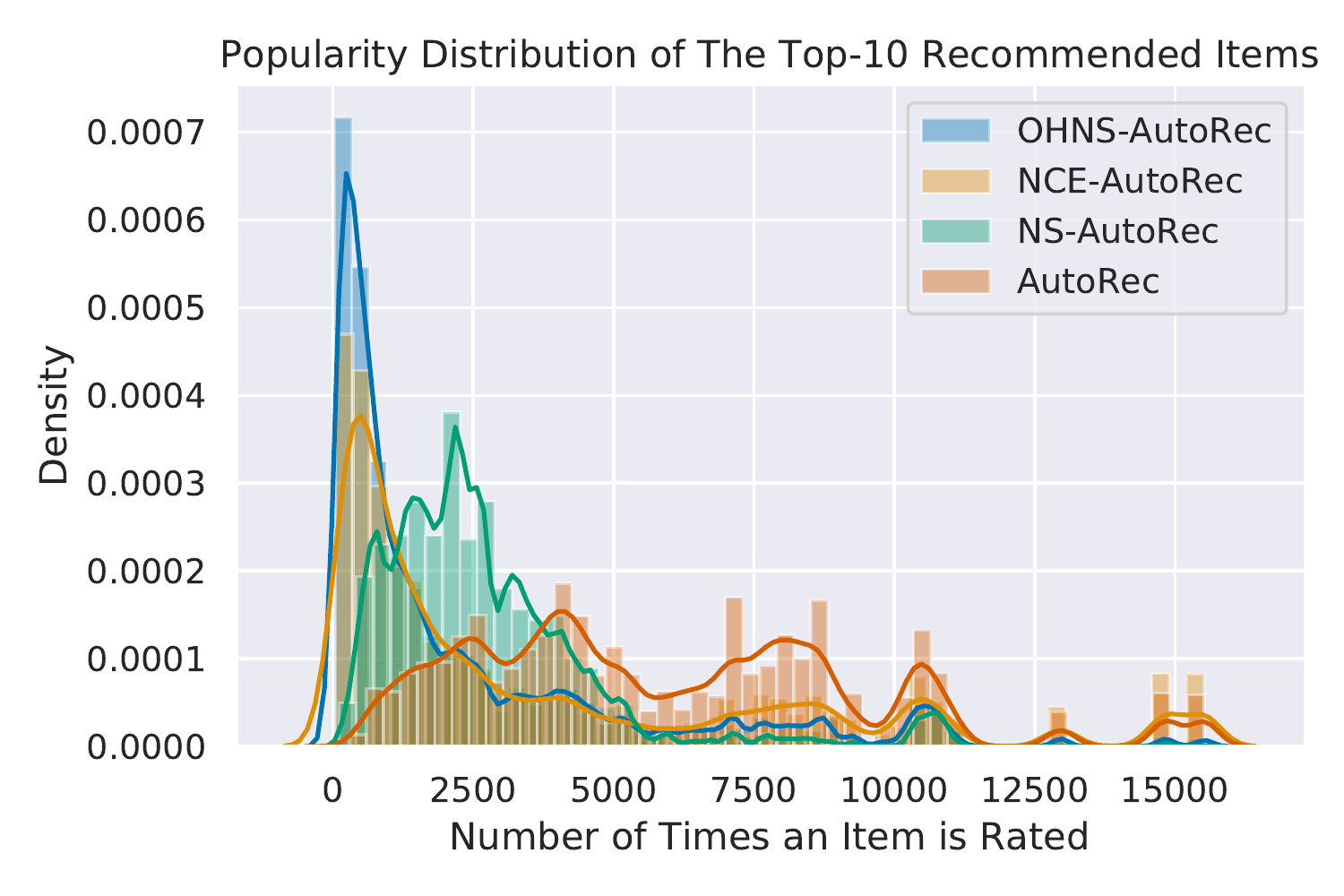}
  \vspace{-1.5mm}
  \caption{Popularity distribution of items recommended for all the users on Goodbooks. 
  A larger value on the x-axis indicates more popularity bias and thus less personalization.}
  \vspace{-1.5mm}
  \label{fig:popularity}
\end{figure}

We now turn to RQ3 concerning the question of popularity-bias and the level of personalization of the different recommenders.
To answer the first part of RQ3, we illustrate personalized recommendation by analyzing the popularity distribution of items recommended by AutoRec, OHNS-AutoRec, NS-AutoRec and NCE-AutoRec. Generally speaking, if an algorithm recommends less popular items, we say it is more personalized. The distribution of items recommended is shown in Figure \ref{fig:popularity}. We observe that OHNS-AutoRec recommends mostly unpopular items since its objective
heavily penalizes popular items. In contrast, AutoRec
exhibits a shift much more to the popular end of the spectrum.  It is worth noting that both NS-AutoRec and NCE-AutoRec focus their recommendations on less popular and hence more nuanced items that indicate a higher level of personalization than the original AutoRec.
Additionally, both NS-AutoRec and NCE-AutoRec outperform AutoRec and other Autoencoder based methods in most evaluations, which suggests that NS-AutoRec and NCE-AutoRec achieve "de-popularization" and benefit from it. Furthermore, it can be seen that for recommending unpopular items, NCE-AutoRec follows OHNS-AutoRec's performance pattern more closely and this might explain why NCE-AutoRec generally outperforms NS-AutoRec.

\subsection{Cold-start Analysis}
\begin{figure}[t]%
\centering
\begin{subfigure}{0.49\linewidth}
\centering
\includegraphics[width=\linewidth]{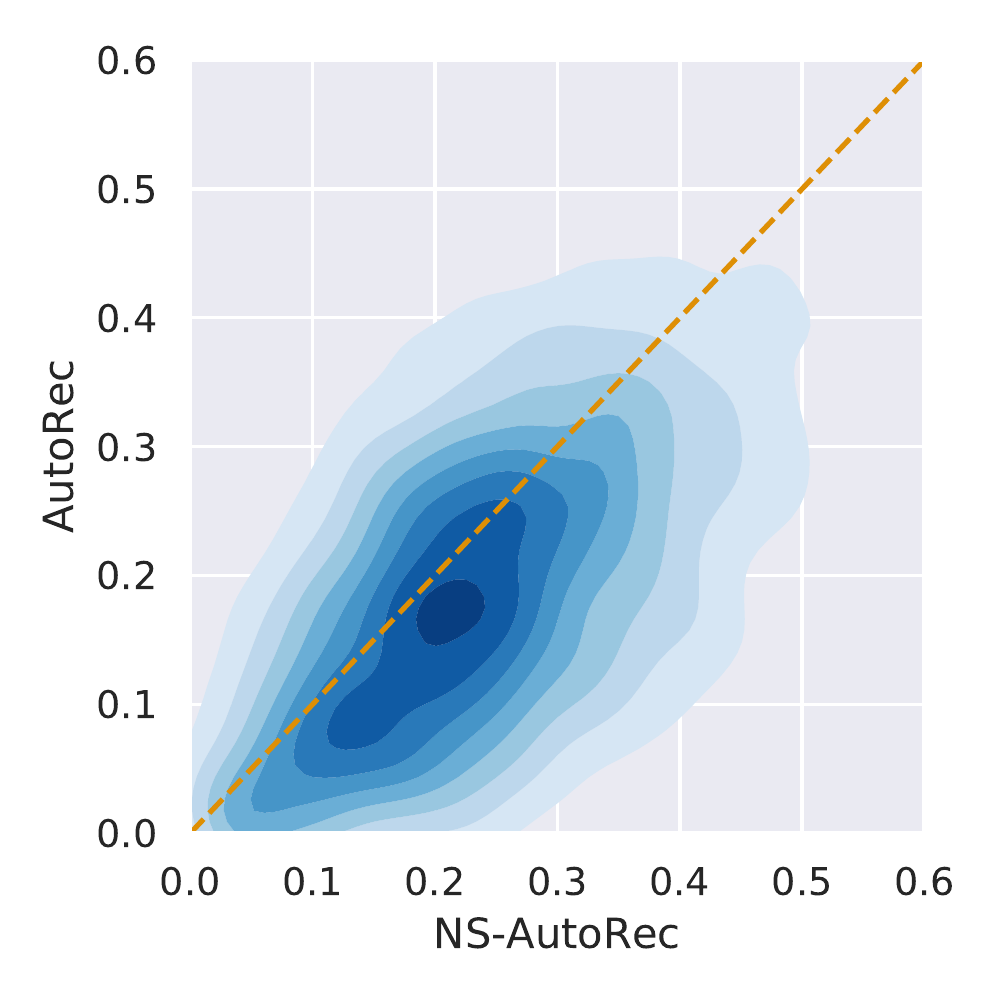}%
\end{subfigure}\hfill%
\begin{subfigure}{0.49\linewidth}
\centering
\includegraphics[width=\linewidth]{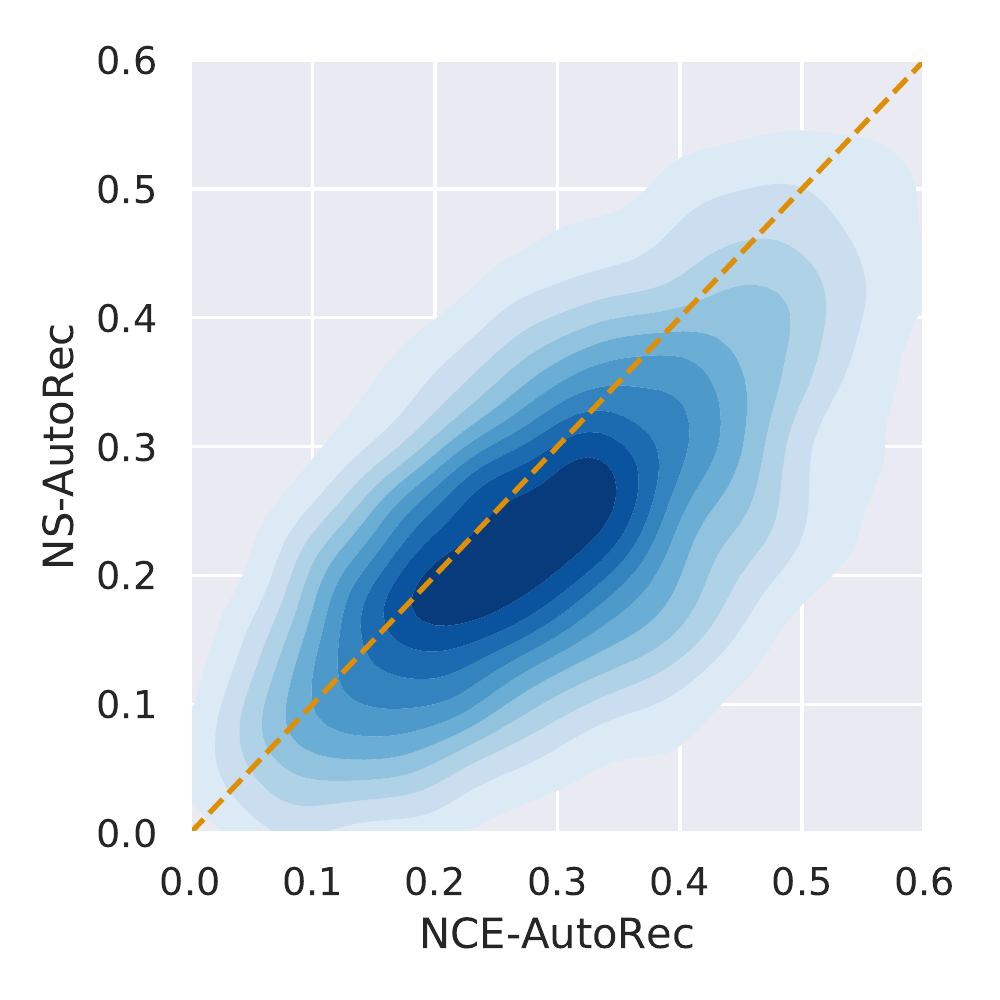}%
\end{subfigure}\hfill%
\caption{Bivariate density plot for pairwise NDCG difference between AutoRec, NS-AutoRec and NCE-AutoRec on Goodbooks. A higher density below line y = x indicates that method portrayed on x-axis has better cold-start user performance.}
\label{fig:cold_start}
\end{figure}

To perform a cold-start user study for RQ3, we hold out 5\% of the users from the training set and carry out training as normal. Then, we evaluate the performance of the model on these users by recommending items to them. We illustrate three pairwise cold-start user evaluation plots on NDCG as shown in Figure \ref{fig:cold_start}. We observe that NCE-AutoRec provides the best recommendation for cold-start users followed by NS-AutoRec then AutoRec.

\subsection{Training Efficiency}
We analyze the training efficiency (RQ4) of NS-AutoRec and NCE-AutoRec with different optimization settings across the four datasets. We plot their corresponding training convergence time on a computer cluster with a 4-core CPU, 32GB RAM and two Nvidia Titan XP GPUs in Figure \ref{fig:nsautorec_nceautorec_time}. Intuitively, we observe that the convergence time increases as the size and sparsity of the dataset increases. Both \textbf{Limited Fine-tune} and \textbf{Full Fine-tune} take about two times as much time as \textbf{Joint} and \textbf{Alternating} because they need to train each head at a separate forward pass, making \textbf{Limited Fine-tune} and \textbf{Full Fine-tune} less efficient in general. By comparing Figure \ref{fig:nsautorec_nceautorec_time_b} with \ref{fig:nsautorec_nceautorec_time_a}, it is worth noting that the training time of these two methods is comparable to each other for smaller datasets, but as the number of users and items become large, NCE-AutoRec converges much faster in terms of training time than NS-AutoRec. This is due to the time consumption of Negative Sampling in NS-AutoRec, which increases drastically as the dataset gets larger in size.

\begin{table*}[t]
\caption{Comparison of recommendation methods on Goodbooks dataset with 95\% confidence interval}
\vspace{-1mm}
\resizebox{\textwidth}{!}{%
\begin{tabular}{c|ccccccccc}
\hline
&model  & R-Precision      & NDCG      & MAP@5      & MAP@50    & Precision@5    & Precision@50    & Recall@5      & Recall@50    \\
\hline
& BPR & 0.2265$\pm$0.0010 & 0.3105$\pm$0.0012 & 0.3851$\pm$0.0027  & 0.2570$\pm$0.0012 & 0.3549$\pm$0.0023 & 0.1872$\pm$0.0008 & 0.0605$\pm$0.0004 & 0.3123$\pm$0.0012 \\
&CDAE & 0.1711$\pm$0.0009 & 0.2427$\pm$0.0012 & 0.3733$\pm$0.0028 & 0.2099$\pm$0.0011 & 0.3280$\pm$0.0023 & 0.1371$\pm$0.0007 & 0.0547$\pm$0.0004 & 0.2229$\pm$0.0010 \\
&CML      & 0.2395$\pm$0.0010 & 0.3285$\pm$0.0013 & 0.4192$\pm$0.0028 & 0.2737$\pm$0.0013 & 0.3856$\pm$0.0024 & 0.1947$\pm$0.0008 & 0.0662$\pm$0.0005 & 0.3259$\pm$0.0012 \\
&NCE-PLRec & 0.2997$\pm$0.0011 & 0.4091$\pm$0.0014 & 0.5789$\pm$0.0028  & 0.3561$\pm$0.0014 & 0.5345$\pm$0.0025 & 0.2315$\pm$0.0009 & 0.0911$\pm$0.0005 & 0.3830$\pm$0.0013 \\
Baselines&PureSVD  & 0.2113$\pm$0.0011 & 0.3013$\pm$0.0012 & 0.4828$\pm$0.0028 & 0.2645$\pm$0.0012 & 0.4263$\pm$0.0025 & 0.1655$\pm$0.0007 & 0.0713$\pm$0.0004 & 0.2698$\pm$0.0010 \\
&VAE-CF & 0.2863$\pm$0.0011 & 0.3884$\pm$0.0014 & 0.5208$\pm$0.0027 & 0.3311$\pm$0.0013 & 0.4824$\pm$0.0024 & 0.2232$\pm$0.0008 & 0.0836$\pm$0.0005 & 0.3729$\pm$0.0013 \\
&WRMF    & 0.2915$\pm$0.0011 & 0.3955$\pm$0.0013 & 0.5314$\pm$0.0027 & 0.3400$\pm$0.0013 & 0.4973$\pm$0.0024 & 0.2278$\pm$0.0008 & 0.0854$\pm$0.0005 & 0.3785$\pm$0.0012 \\
&AutoRec & 0.2069$\pm$0.0009 & 0.2927$\pm$0.0012 & 0.4601$\pm$0.0028 & 0.2554$\pm$0.0012 & 0.4048$\pm$0.0024 & 0.1621$\pm$0.0007 & 0.0681$\pm$0.0004 & 0.2643$\pm$0.0011 \\
\hline
&OHNS-AutoRec & 0.1918$\pm$0.0009 & 0.2787$\pm$0.0012 & 0.4628$\pm$0.0028 & 0.2423$\pm$0.0012 & 0.4007$\pm$0.0024 & 0.1475$\pm$0.0007 & 0.0685$\pm$0.0005 & 0.2470$\pm$0.0011 \\
Proposed Models&NS-AutoRec      & 0.2516$\pm$0.0010 & 0.3493$\pm$0.0013 & 0.5102$\pm$0.0028 & 0.3018$\pm$0.0013 & 0.4599$\pm$0.0024 & 0.1975$\pm$0.0008 & 0.0779$\pm$0.0004 & 0.3260$\pm$0.0011 \\
&NCE-AutoRec      & \textbf{0.3090$\pm$0.0012} & \textbf{0.4215$\pm$0.0014} & \textbf{0.5830$\pm$0.0028} & \textbf{0.3627$\pm$0.0014} & \textbf{0.5355$\pm$0.0025} & \textbf{0.2407$\pm$0.0009} & \textbf{0.0916$\pm$0.0005} & \textbf{0.4001$\pm$0.0013} \label{table:goodbooks} \\
\hline
\end{tabular}
}
\end{table*}

\begin{table*}[t]
\caption{Comparison of recommendation methods on MovieLens-20M dataset with 95\% confidence interval}
\vspace{-1mm}
\resizebox{\textwidth}{!}{%
\begin{tabular}{c|ccccccccc}
\hline
&model  & R-Precision      & NDCG      & MAP@5      & MAP@50    & Precision@5    & Precision@50    & Recall@5      & Recall@50      \\
\hline
&BPR & 0.0844$\pm$0.0006 & 0.1540$\pm$0.0008 & 0.1188$\pm$0.0012  & 0.0917$\pm$0.0006 & 0.1133$\pm$0.0010 & 0.0760$\pm$0.0005 & 0.0403$\pm$0.0005 & 0.2274$\pm$0.0013 \\
&CDAE & 0.0840$\pm$0.0006 & 0.1538$\pm$0.0008 & 0.1260$\pm$0.0012 & 0.0933$\pm$0.0006 & 0.1176$\pm$0.0010 & 0.0757$\pm$0.0005 & 0.0402$\pm$0.0005 & 0.2226$\pm$0.0013 \\
&CML      & 0.0890$\pm$0.0006 & 0.1771$\pm$0.0008 & 0.1228$\pm$0.0011 & 0.1019$\pm$0.0006 & 0.1188$\pm$0.0010 & 0.0873$\pm$0.0005 & 0.0388$\pm$0.0005 & 0.2799$\pm$0.0014 \\
&NCE-PLRec & 0.1020$\pm$0.0006 & 0.1957$\pm$0.0009 & 0.1456$\pm$0.0012 & 0.1113$\pm$0.0006 & 0.1370$\pm$0.0010 & 0.0917$\pm$0.0005 & 0.0498$\pm$0.0005 & 0.2971$\pm$0.0014 \\
Baselines&PureSVD  & 0.0954$\pm$0.0006 & 0.1783$\pm$0.0008 & 0.1375$\pm$0.0012 & 0.1041$\pm$0.0006 & 0.1286$\pm$0.0010 & 0.0850$\pm$0.0004 & 0.0451$\pm$0.0005 & 0.2647$\pm$0.0012 \\
&VAE-CF & 0.1005$\pm$0.0006 & 0.1972$\pm$0.0009 & 0.1363$\pm$0.0012 & 0.1072$\pm$0.0006 & 0.1296$\pm$0.0010 & 0.0898$\pm$0.0004 & 0.0502$\pm$0.0005 & 0.3073$\pm$0.0014 \\
&WRMF    & 0.1000$\pm$0.0006 & 0.1962$\pm$0.0008 & 0.1433$\pm$0.0012 & 0.1106$\pm$0.0006 & 0.1342$\pm$0.0010 & 0.0918$\pm$0.0004 & 0.0485$\pm$0.0006 & 0.3021$\pm$0.0014 \\
&AutoRec & 0.0937$\pm$0.0006 & 0.1704$\pm$0.0009 & 0.1401$\pm$0.0012 & 0.1022$\pm$0.0006 & 0.1298$\pm$0.0010 & 0.0824$\pm$0.0005 & 0.0457$\pm$0.0005 & 0.2458$\pm$0.0013 \\
\hline
&OHNS-AutoRec & 0.0783$\pm$0.0005 & 0.1461$\pm$0.0007 & 0.1202$\pm$0.0011 & 0.0851$\pm$0.0005 & 0.1107$\pm$0.0009 & 0.0650$\pm$0.0003 & 0.0400$\pm$0.0005 & 0.2124$\pm$0.0011 \\
Proposed Models&NS-AutoRec      & 0.1026$\pm$0.0006 & 0.1975$\pm$0.0008 & 0.1465$\pm$0.0012 & 0.1120$\pm$0.0006 & 0.1378$\pm$0.0010 & 0.0921$\pm$0.0004 & \textbf{0.0508$\pm$0.0005} & 0.3003$\pm$0.0014 \\
&NCE-AutoRec      & \textbf{0.1034$\pm$0.0006} & \textbf{0.2012$\pm$0.0009} & \textbf{0.1488$\pm$0.0013} & \textbf{0.1141$\pm$0.0006} & \textbf{0.1395$\pm$0.0011} & \textbf{0.0943$\pm$0.0005} & 0.0498$\pm$0.0005 & \textbf{0.3079$\pm$0.0013} \label{table:ml-20m} \\
\hline
\end{tabular}
}
\end{table*}

\begin{table*}[t]
\caption{Comparison of recommendation methods on Netflix dataset with 95\% confidence interval}
\vspace{-1mm}
\resizebox{\textwidth}{!}{%
\begin{tabular}{c|ccccccccc}
\hline
&model  & R-Precision      & NDCG      & MAP@5      & MAP@50    & Precision@5    & Precision@50    & Recall@5      & Recall@50     \\
\hline
&BPR & 0.0772$\pm$0.0003 & 0.1307$\pm$0.0004 & 0.1154$\pm$0.0006  & 0.0956$\pm$0.0003 & 0.1130$\pm$0.0005 & 0.0814$\pm$0.0002 & 0.0306$\pm$0.0002 & 0.1844$\pm$0.0006 \\
&CDAE & 0.0800$\pm$0.0003 & 0.1321$\pm$0.0004 & 0.1252$\pm$0.0006 & 0.0985$\pm$0.0003 & 0.1208$\pm$0.0005 & 0.0836$\pm$0.0002 & 0.0323$\pm$0.0002 & 0.1800$\pm$0.0006 \\
&CML      & 0.0882$\pm$0.0003 & 0.1515$\pm$0.0004 & 0.1406$\pm$0.0006 & 0.1094$\pm$0.0003 & 0.1342$\pm$0.0005 & 0.0907$\pm$0.0003 & 0.0369$\pm$0.0002 & 0.2120$\pm$0.0006 \\
&NCE-PLRec & 0.1049$\pm$0.0003 & 0.1764$\pm$0.0004 & 0.1654$\pm$0.0007 & 0.1247$\pm$0.0003 & 0.1552$\pm$0.0006 & 0.1015$\pm$0.0003 & 0.0477$\pm$0.0003 & 0.2392$\pm$0.0007 \\
Baselines&PureSVD & 0.0994$\pm$0.0003 & 0.1644$\pm$0.0004 & 0.1590$\pm$0.0007 & 0.1180$\pm$0.0003 & 0.1490$\pm$0.0005 & 0.0953$\pm$0.0003 & 0.0445$\pm$0.0003 & 0.2188$\pm$0.0006 \\
&VAE-CF & 0.1009$\pm$0.0003 & 0.1706$\pm$0.0004 & 0.1520$\pm$0.0006 & 0.1164$\pm$0.0003 & 0.1431$\pm$0.0005 & 0.0954$\pm$0.0003 & 0.0464$\pm$0.0003 & 0.2336$\pm$0.0006 \\
&WRMF    & 0.0985$\pm$0.0003 & 0.1681$\pm$0.0004 & 0.1531$\pm$0.0007 & 0.1170$\pm$0.0003 & 0.1447$\pm$0.0006 & 0.0960$\pm$0.0003 & 0.0450$\pm$0.0003 & 0.2325$\pm$0.0007 \\
&AutoRec & 0.0903$\pm$0.0003 & 0.1504$\pm$0.0004 & 0.1425$\pm$0.0006 & 0.1098$\pm$0.0003 & 0.1344$\pm$0.0005 & 0.0912$\pm$0.0003 & 0.0377$\pm$0.0003 & 0.2030$\pm$0.0006 \\
\hline
&OHNS-AutoRec & 0.0811$\pm$0.0003 & 0.1256$\pm$0.0004 & 0.1437$\pm$0.0007 & 0.0987$\pm$0.0003 & 0.1328$\pm$0.0005 & 0.0742$\pm$0.0003 & 0.0361$\pm$0.0003 & 0.1491$\pm$0.0006 \\
Proposed Models&NS-AutoRec      & 0.0993$\pm$0.0003 & 0.1646$\pm$0.0004 & 0.1581$\pm$0.0007 & 0.1176$\pm$0.0003 & 0.1483$\pm$0.0005 & 0.0952$\pm$0.0003 & 0.0445$\pm$0.0003 & 0.2199$\pm$0.0006 \\
&NCE-AutoRec      & \textbf{0.1078$\pm$0.0003} & \textbf{0.1797$\pm$0.0004} & \textbf{0.1684$\pm$0.0007} & \textbf{0.1265$\pm$0.0003} & \textbf{0.1582$\pm$0.0006} & \textbf{0.1026$\pm$0.0003} & \textbf{0.0494$\pm$0.0003} & \textbf{0.2427$\pm$0.0007} \label{table:netflix} \\
\hline
\end{tabular}
}
\end{table*}

\begin{table*}[t]
\caption{Comparison of recommendation methods on Yahoo dataset with 95\% confidence interval}
\vspace{-1mm}
\resizebox{\textwidth}{!}{%
\begin{tabular}{c|ccccccccc}
\hline
&model  & R-Precision      & NDCG      & MAP@5      & MAP@50    & Precision@5    & Precision@50    & Recall@5      & Recall@50     \\
\hline
&BPR & 0.1127$\pm$0.0002 & 0.2422$\pm$0.0003 & 0.1335$\pm$0.0003  & 0.0940$\pm$0.0002 & 0.1267$\pm$0.0003 & 0.0674$\pm$0.0001 & 0.0920$\pm$0.0002 & 0.4236$\pm$0.0004 \\
&CDAE & 0.0795$\pm$0.0002 & 0.1844$\pm$0.0002 & 0.1013$\pm$0.0001 & 0.0671$\pm$0.0002 & 0.0889$\pm$0.0002 & 0.0509$\pm$0.0001 & 0.0663$\pm$0.0002 & 0.3327$\pm$0.0004 \\
&CML      & 0.1862$\pm$0.0003 & 0.3723$\pm$0.0003 & 0.2413$\pm$0.0002 & 0.1488$\pm$0.0004 & 0.2203$\pm$0.0002 & 0.0983$\pm$0.0001 & 0.1549$\pm$0.0003 & 0.6095$\pm$0.0004 \\
&NCE-PLRec & 0.2530$\pm$0.0003 & 0.4555$\pm$0.0004 & 0.3410$\pm$0.0002 & 0.1832$\pm$0.0004 & 0.2964$\pm$0.0002 & 0.1097$\pm$0.0002 & 0.2199$\pm$0.0004 & 0.6722$\pm$0.0004 \\
Baselines&PureSVD  & 0.2041$\pm$0.0003 & 0.3662$\pm$0.0003 & 0.2829$\pm$0.0002 & 0.1491$\pm$0.0004 & 0.2447$\pm$0.0002 & 0.0889$\pm$0.0001 & 0.1770$\pm$0.0003 & 0.5352$\pm$0.0004 \\
&VAE-CF & 0.2515$\pm$0.0003 & 0.4462$\pm$0.0004 & 0.3252$\pm$0.0002 & 0.1745$\pm$0.0004 & 0.2837$\pm$0.0002 & 0.1043$\pm$0.0001 & 0.2203$\pm$0.0004 & 0.6563$\pm$0.0004 \\
&WRMF    & 0.2351$\pm$0.0003 & 0.4258$\pm$0.0003 & 0.2866$\pm$0.0002 & 0.1614$\pm$0.0003 & 0.2540$\pm$0.0002 & 0.1005$\pm$0.0001 & 0.2116$\pm$0.0004 & 0.6445$\pm$0.0004 \\
&AutoRec & 0.1205$\pm$0.0002 & 0.2498$\pm$0.0003 & 0.1679$\pm$0.0002 & 0.0997$\pm$0.0003 & 0.1458$\pm$0.0003 & 0.0666$\pm$0.0001 & 0.0979$\pm$0.0002 & 0.4065$\pm$0.0004 \\
\hline
&OHNS-AutoRec & 0.1912$\pm$0.0003 & 0.3097$\pm$0.0004 & 0.2736$\pm$0.0002 & 0.1346$\pm$0.0004 & 0.2346$\pm$0.0002 & 0.0728$\pm$0.0002 & 0.1633$\pm$0.0004 & 0.3974$\pm$0.0004 \\
Proposed Models&NS-AutoRec      & 0.2626$\pm$0.0003 & 0.4614$\pm$0.0004 & 0.3525$\pm$0.0002 & 0.1835$\pm$0.0004 & 0.3050$\pm$0.0002 & 0.1065$\pm$0.0002 & 0.2346$\pm$0.0004 & 0.6653$\pm$0.0004 \\
&NCE-AutoRec      & \textbf{0.2681$\pm$0.0003} & \textbf{0.4728$\pm$0.0004} & \textbf{0.3575$\pm$0.0002} & \textbf{0.1890$\pm$0.0004} & \textbf{0.3107$\pm$0.0002} & \textbf{0.1108$\pm$0.0001} & \textbf{0.2380$\pm$0.0004} & \textbf{0.6858$\pm$0.0004} \label{table:yahoo} \\
\hline
\end{tabular}
}
\end{table*}

\section{Conclusion}
In this work, we presented two novel two-headed Autoencoder based recommendation algorithms called NS-AutoRec and NCE-AutoRec that respectively use NS and NCE optimization objectives via one head to learn embeddings that mitigate the strong popularity bias that we observed in AutoRec. For the other head, we employ the standard AutoRec MSE reconstruction objective 
for end task 
recommendation predictions. 

On four real-world large-scale OC-CF datasets, our contributions of NCE-AutoRec and NS-AutoRec showed that they achieve competitive performance in comparison to  state-of-the-art recommenders including BPR, AutoRec, CDAE, and VAE-CF.  Furthemore, critical to the the goal of this paper to ``de-popularize'' AutoRec, NCE-AutoRec and NS-AutoRec recommend more nuanced (i.e., less overall popular) items than AutoRec that lend them a greater degree of personalization for end users.  Comparing NS-AutoRec and NCE-AutoRec, we observe that NCE-AutoRec often slightly outperforms the NS variant and typically trains to convergence in an order of magnitude less time. 
Overall, this work has shown how recent advances leveraging NS and NCE to train high-quality embeddings can be extended to Autoencoder-based recommenders, thus yielding novel enhancements of these methods and state-of-the-art \emph{highly personalized} OC-CF recommendation performance.  These encouraging results should open the door to future NS and NCE two-headed adaptations of other embedding-based recommendation methods where the level of personalization can be improved.
\newpage
\bibliographystyle{unsrt}
\bibliography{sample-base}

\end{document}